\documentclass[12pt]{article}
\usepackage[margin=0.95 in]{geometry}
\usepackage{amsmath}
\usepackage{amssymb,amsfonts}
\usepackage[all]{xy}
\usepackage{graphicx}
\usepackage[utf8x]{inputenc}
\usepackage{amsmath}
\usepackage{amssymb}
\usepackage{float}
\usepackage{array}
\usepackage{tikz,caption}
\usetikzlibrary{shapes.geometric, calc,positioning}
\usepackage{mathtools}
\usepackage{mathrsfs}
\usepackage{comment}
\usepackage{hyperref}
\usepackage{cite}
\usepackage{dirtytalk}
\numberwithin{equation}{section}
\numberwithin{equation}{section}
\numberwithin{table}{section}

   \definecolor{newblue}{rgb}{0,0,.6}
\definecolor{newred}{RGB}{178,34,34}
\definecolor{newgreen}{rgb}{0.18, 0.65, 0.3}
\definecolor{newgray}{RGB}{101,101,101}
\definecolor{newyellow}{rgb}{.96,.78,.10}
\definecolor{newpurple}{RGB}{128,0,128}
\definecolor{newteal}{RGB}{0,128,128}
\definecolor{newpink}{rgb}{1,.49,.47}

\newcommand{\be}{\begin{equation}}
\newcommand{\ee}{\end{equation}}

\begin{document}
\begin{flushright}
KEK-TH-2240
\end{flushright}
\thispagestyle{empty}

\vspace*{3cm}
{}
\noindent
{\Large \bf  
Weights, Recursion relations and Projective triangulations for positive geometry of scalar theories}
\vskip .4cm
\noindent
\linethickness{.06cm}
\line(10,0){467}
\vskip 1.1cm
\noindent
\noindent
{\large \bf Renjan Rajan John$^a$, Ryota Kojima$^b$, Sujoy Mahato$^c$}
\vskip 0.5cm
 
\noindent
$^a$Universit\`a del Piemonte Orientale, Dipartimento di Scienze e Innovazione Tecnologica\\
Viale T. Michel 11, I-15121 Alessandria, Italy
 \vskip 0.5cm

\noindent
$^a$I.\,N.\,F.\,N. - sezione di Torino\\
 Via P. Giuria 1, I-10125 Torino, Italy
 \vskip 0.5cm
 
 \noindent
$^b$KEK Theory Center\\
 Tsukuba, Ibaraki, 305-0801, Japan
 \vskip 0.5cm

 \noindent
$^c$Institute of Mathematical Sciences \\
Homi Bhabha National Institute (HBNI)\\
IV Cross Road, C.~I.~T.~Campus, \\
  Taramani, Chennai, 600113  Tamil Nadu, India
 \vskip 0.5cm
 \noindent
renjan.rajan@to.infn.it, ryota@post.kek.jp, sujoymahato@imsc.res.in 

\vskip 1.1cm


\noindent {\sc Abstract: }  
The story of positive geometry of massless scalar theories was pioneered in \cite{ABHY} in the context of bi-adjoint $\phi^3$ theories. Further study proposed that the positive geometry for a generic massless scalar theory with polynomial interaction is a class of polytopes called accordiohedra \cite{Mrunmay}. Tree-level planar scattering amplitudes of the theory can be obtained from a weighted sum of the canonical forms of the accordiohedra. In this paper, using results of the recent work \cite{Ryota}, we show that in theories with polynomial interactions all the weights can be determined from the factorization property of the accordiohedron. We also extend the projective recursion relations introduced in \cite{CausalDiamonds,Yang} to these theories. We then give a detailed analysis of how the recursion relations in $\phi^p$ theories and theories with polynomial interaction correspond to projective triangulations of accordiohedra. Following the very recent development \cite{alok1loop} we also extend our analysis to one-loop integrands in the quartic theory.
\vskip1cm


\newpage
\setcounter{tocdepth}{2}
\tableofcontents
\section{Introduction}
In recent years, tremendous progress has been made in relating scattering amplitudes to interesting mathematical and geometric structures. One of the issues with this progress was that the relevant geometry lived in an auxiliary space as opposed to the kinematic space where the S-matrix lives. This was overcome in \cite{ABHY} where the tree-level amplitudes of the bi-adjoint $\phi^3$ theory \cite{CHY} were related to canonical forms of a positive geometry called the kinematic space associahedron \footnote{For a recent review on related developments see \cite{Review}.}. The program was extended to the case of planar tree-level amplitudes of the $\phi^4$ theory in \cite{Alok}. There it was shown that the relevant geometry in kinematic space for the quartic theory was the Stokes polytope \cite{Stokes1,Stokes2}. The positive geometry of $\phi^p$ theories $(p>4)$ and theories with polynomial interactions (such as $\phi^3+\phi^4$) was studied in \cite{Raman} and \cite{Mrunmay} respectively. In these works, the positive geometry associated to a general class of scalar theories was understood to be the accordiohedron and the explicit convex embedding of a few lower dimensional polytopes in the kinematic space was given. A canonical convex embedding of these accordiohedra in the kinematic space was provided in \cite{1906ppp}. For an algorithmic procedure to obtain such embeddings see \cite{Ours}. For other interesting related developments, we refer the reader to \cite{Mizera:2017cqs} in the context of string KLT relations, \cite{Mizera:2017rqa,Kalyanapuram:2019nnf,Kalyanapuram:2020vil} in the context of intersection theory.

\par

Unlike in the cubic theory, in theories with $\phi^p$ or polynomial interactions, the geometry related to each
amplitude is not a single accordiohedron, but a collection of them. The weighted sum of the canonical forms of these polytopes
gives the complete amplitude. It is an important problem whether we need additional data of the quantum field theory to determine amplitudes from the geometry. In this paper, we will show that as in $\phi^p$ theories \cite{Ryota}, all weights can be determined from the factorization property of the positive geometry even in theories with polynomial interaction. 
This means that, for general scalar theory, it is possible to obtain the
scattering amplitude purely from the positive geometry. 

Using geometric properties of the associahedron and the corresponding canonical forms, recursion relations for these forms (and thereby for scattering amplitudes in $\phi^3$ theory) were obtained in \cite{ABHY}. A purely field theoretic derivation of these relations was obtained in \cite{Song-Yang} by introducing a one complex variable rescaling of all the basis planar kinematic variables $X_{ij}\rightarrow z\,X_{ij}$, and constructing a meromorphic function in the complex plane without a residue at $z=0$ or $z=\infty$.  Furthermore \cite{Song-Yang} provided solutions to these recursion relations for the general case. Each term in the recursion corresponds to the canonical function of one of the simplices obtained by the full triangulation of the associahedron, reminiscent of the BCFW representation of $\mathcal N=4$ sYM amplitudes. This was recently studied for $\phi^p$ theories in \cite{Ryota}. 

In \cite{CausalDiamonds} new recursion relations for the cubic theory were introduced by projecting the associahedron onto one of its facets which then divides the associahedron into prisms. This corresponds to what is called the projective triangulation of the associahedron and it has the interesting feature that it divides the original geometry into positive geometries that can have curvy facets.
The recursion relations corresponding to such triangulations were derived from field theory in \cite{Yang}  for tree and one-loop amplitudes of the bi-adjoint $\phi^3$ theory. In this paper, we extend the study of projective triangulations to generic scalar field theories. 
We make use of the canonical convex embedding of accordiohedra in kinematic space to show that the recursion relations of \cite{Song-Yang,Yang} hold for such theories. We then relate the representation of amplitudes via. such recursion relations to projective triangulations of the relevant accordiohedron. We also extend our analysis to one-loop integrands in the quartic theory using a simple example, following the recent work \cite{alok1loop}.

The paper is organized as follows. In Section \ref{review} we review some aspects of the accordiohedron relevant for the rest of the paper. In Section \ref{weights} we compute weights of  accordiohedra that contribute to the planar amplitude of various scalar theories. In Section \ref{derivation} following \cite{Yang} we give a proof of the recursion relations that we work with. In Section \ref{Examples} we make use of the canonical convex embedding of accordiohedra in kinematic space and the recursion relations to compute the amplitudes in a few explicit examples.
In Section \ref{triangulations} we give the geometric interpretation of our results from recursion in terms of projective triangulations of accordiohedra. In Section \ref{loopsection} we consider one-loop integrands of the quartic theory and illustrate the recursion relations and their relation to projective triangulations of polytopes in a simple example. We end with a conclusion in Section \ref{conclusion} and an example in Appendix \ref{App}.

\section{Review of the accordiohedron}
\label{review}
In this section, we briefly review the accordiohedron which was proposed as the positive geometry for scalar field theories with $\phi^p$ interactions in \cite{Raman} and polynomial interactions in \cite{Mrunmay}. We will closely follow the discussion in \cite{Raman,Mrunmay}.
\subsection{$D$-accordiohedron and the canonical form}
Here we define a $D$-accordiohedron which was proposed as the positive geometry for  scalar field theories with polynomial interactions in \cite{Mrunmay}. First, we consider two $n$-point convex polygons: a polygon $P$ and another one $P'$ whose vertices $i'$ are the mid-points of the edge connecting the vertices $i$ and $i+1$ of $P$.
\begin{figure}[H]
\centering
\includegraphics[width=140mm]{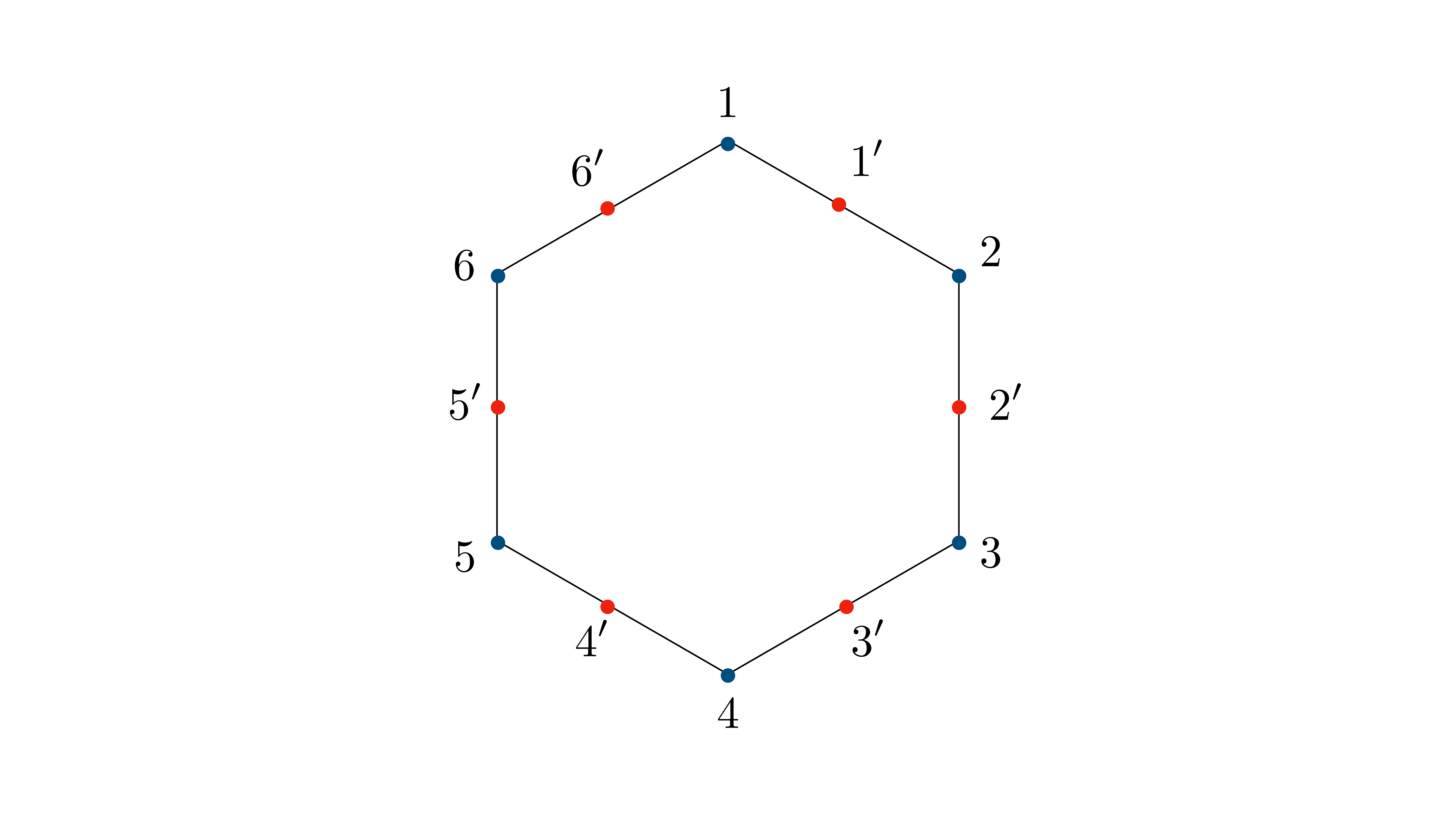}
\caption{Two polygons: $P$ and $P'$.}
\label{figpolygon}
\end{figure}
We denote a dissection of $P$ as $D$.  We define a cut $C((i'j'),D)$ of the diagonal $(i',j')$ of $P'$ as a set of the lines $(i,i+1)$, $(j,j+1)$ of $P$ and the diagonals of the $D$ which intersect the diagonal $(i',j')$. If the cut $C((i'j'),D)$ is connected, we say that the diagonal $(i',j')$ of $P'$ is ``compatible" with the dissection $D$ of $P$.
\begin{figure}[H]
\centering
\includegraphics[width=140mm]{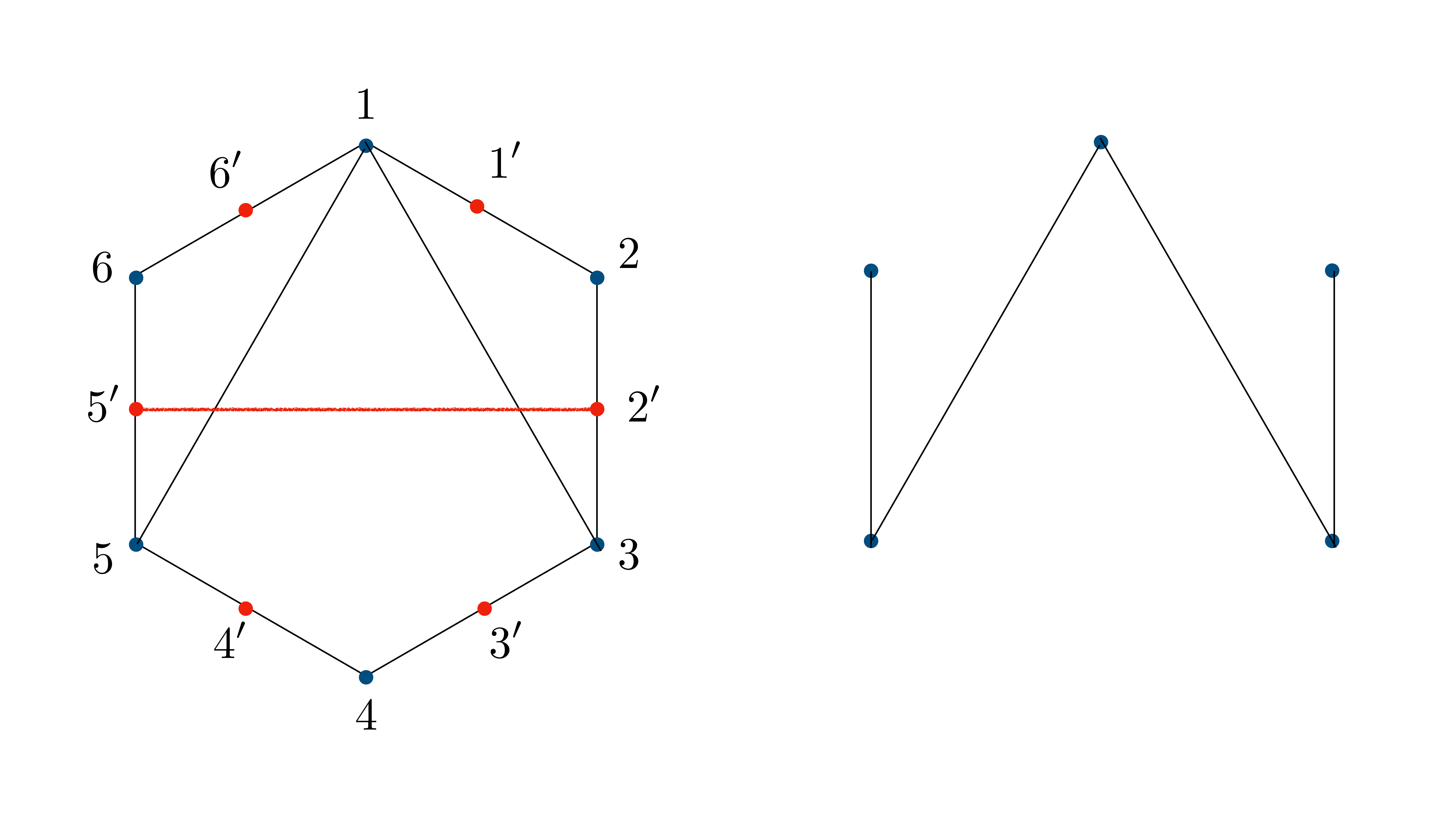}
\caption{Dissection $\{(1,3),(1,5)\}$ on the left and the cut by $(2',5')$ on the right.}
\label{figconnected}
\end{figure}
\par
A dissection of $P'$ consisting of diagonals compatible with the dissection $D$ is called ``$D$-accordion dissection". Then the $D$-accordiohedron $\mathcal{AC}(D)$ is defined as a polytope whose vertices are the $D$-accordion dissections. \par
If we restrict the dissection $D$ to all possible $p$-angulations, the $D$-accordiohedron corresponds to the accordiohedron. The notion of compatibility defined above corresponds to the $Q$-compatibility of the $p$-angulations. Then the $D$-accordiohedron defined above can be interpreted as the generalization of the accordiohedron.\par
Since the $D$-accordiohedron is a simple polytope, the canonical form is a sum over its vertices. We denote each vertex as $Z$ and the diagonal $(i,j)$ of $P$ as $X_{ij}$. We also label the facets which adjacent with the vertex $Z$ as $X_{i_a,j_a}=0$. The canonical form is given by \cite{ABHY,Arkani-Hamed:2017tmz,Salvatori:2019phs}:
\begin{align}
\Omega(\mathcal{AC}(D))=\sum_{\text{vertex}\ Z}\text{sign}(Z)\bigwedge_{a\in Z}d\log{X_{i_a,j_a}}.
\end{align}
The sign$(Z)$ is fixed by requiring projectivity of the form. \par
\subsection{Planar scattering form and amplitudes}
Planar kinematic variables $X_{i,j}$ are defined as
\begin{align}
X_{i,j}=(p_i+p_{i+1}+\dots +p_j)^2=s_{i,i+1,\dots,j-1}\ \ \text{for}\ \ 1\leq i <j \leq n
\end{align}
The on-shell condition $p_i^2=0$ is translated into $X_{i,i+1}=X_{1,n}=0$. The Mandelstam variables are expressed in terms of the planar variables as :
\begin{align}
s_{ij}=X_{i,j+1}+X_{i+1,j}-X_{i,j}-X_{i+1,j+1}.
\end{align}
We can associate to each planar graph $g$ with propagators $X_{i_1,j_1},\dots,X_{i_n,j_n}$ a form:
\begin{align}
\text{sign}(g)\bigwedge_{a=1}^nd\log{X_{i_a,j_a}}.
\end{align}
There is a one-to-one correspondence between the dissection of the $n$-point polygon with diagonals $(i_1,j_1),(i_2,j_2),\dots,(i_k,j_k)$ and the planar graphs $g$ with propagators $X_{i_1,j_1},X_{i_2,j_2},\dots,X_{i_k,j_k}$. We choose a reference dissection $D$ and consider a set of dissections compatible with this reference dissection. Equivalently, these dissections make a $D$-accordiohedron. Using the correspondence between dissections and graphs, we can obtain a set of graphs from the set of dissections. We define a dissection $D$-dependent planar scattering form $\Omega^D_d$ as
\begin{align}
\Omega^D_d=\sum_{g(D)}\text{sign}(g(D))\bigwedge_{a=1}^dd\log{X_{i_a,j_a}},
\end{align}
where $\text{sign}(g(D))=\pm1$, $d$ is the dimension of the $D$-accordiohedron  and we sum over all planar graphs which are related to all the vertices of the $D$-accordiohedron. The sign is determined by projectivity which is ensured by the following rule :
\begin{align}
\text{sign}(g(D))=-\text{sign}(g'(D))
\end{align}
 where the corresponding dissections of $g$ and $g'$ are adjacent vertices of the $D$-accordiohedron. \par
When we embed the accordiohedron in the kinematic space, its canonical form is the pullback of the planar scattering form onto the accordiohedron. In \cite{Alok, Raman}, the authors used specific embeddings of accordiohedra in kinematic space associahedra by imposing certain additional constraints. In this paper, we use the canonical convex realization of accordiohedra in kinematic space following \cite{1906ppp,Ours}. 
\par
To obtain the scattering amplitude, we need to consider a set of reference dissections $\{D_1,\dots,D_k\}$ called primitive dissections for which:
\begin{itemize}
\item no two dissections of this set are related to each other by cyclic permutations
\item all the other dissections are obtained by a cyclic permutations of dissection
of this subset.
\end{itemize}
Then the scattering amplitude is given as
\begin{align}
\mathcal{M}_n=\sum_{\text{perm}\ \sigma}\sum_{\text{primitive}\ D}\alpha^{D}_nm^{\sigma, D}_n.
\end{align}
where $\alpha^{D}_n$ are the weights and $m^{\sigma, D}_n$ is the ratio part of the scattering form:
\begin{align}
m^{\sigma, D}_n=\sum_{g(D)}\prod_{a=1}^{d}\frac{1}{X_{i_a,j_a}}.
\end{align}
The canonical forms of the accordiohedra obtained from the reference dissection of an $n$-point polygon into $a_3$ triangles, $a_4$ squares, and so on, give graphs which have $a_3$ 3-point vertices, $a_4$ 4-point vertices, and so on. To get the full amplitude, we need to consider all possible dissections.  \par
Let us now consider an example: the amplitude for the $n=5$ case in a theory with $\phi^3+\phi^4$ interaction. We choose the reference dissection $\{1,3\}$ of the pentagon. The  compatible disssection is $\{2,5\}$ and the accordiohedron is a line segment. We can embed this accordiohedron in kinematic space  using the following constraint from the convex realization: 
\begin{align}
X_{13}&=-X_{25}+\epsilon_{13}.
\end{align}
Then its canonical form is given as
\begin{align}
\Omega(\mathcal{AC}^5(13))=dX_{13}\left(\frac{1}{X_{13}}+\frac{1}{X_{25}}\right)=dX_{13}\mathcal{M}_{(13)}.
\end{align}
Similarly we can obtain other canonical forms by cyclic permutations. The weighted sum of these forms is
\begin{align}
\widetilde{\mathcal{M}}_5=\alpha^{(13)}_5\left(\mathcal{M}_{(13)}+\mathcal{M}_{(24)}+\dots +\mathcal{M}_{(52)}\right).
\end{align}
The weight is determined by demanding that the residue in each channel is one. In this case the weight is uniquely fixed as $\alpha^{(13)}_5=\frac{1}{2}$ and $\widetilde{\mathcal{M}}_5$ corresponds to the 5-point amplitude. The important point is that this condition relied on the form of scattering amplitudes. In the next section, we will see that the weights can be determined purely from the geometry.

\section{Weights for the mixed vertices}
\label{weights}
In this section, we determine the weights for various mixed vertices cases from the accordiohedron. 
\subsection{Weights and the factorization}
Here we briefly review the determination of weights from factorization following \cite{Ryota}. For details of this factorization property see \cite{Mrunmay}. First we choose a diagonal $(ij)$ and consider all complete dissections (including those obtained by cyclic permutations of primitive dissections) $P$ which contain this diagonal. From these dissections, we can obtain the accordiohedra $\mathcal{AC}_{n}^{P}$.  The facet $X_{ij}=0$ of this accordiohedron is given by a product of lower dimensional accordiohedra
\begin{equation}
\label{eq:factorizationaccordiohedron}
\mathcal{AC}_{n}^{P}|_{X_{ij}=0}=\mathcal{AC}_{m}^{P_1}\times\mathcal{AC}_{n+2-m}^{P_2},
\end{equation} 
where $P_1$ is the dissection of the polygon $\{i,i+1,\dots,j\}$ and $P_2$ is the dissection of $\{j,j+1,\dots,n,1,\dots,i\}$. These dissection satisfy $P_1\cup P_2\cup (ij)=P$.
This factorization implies physical factorization of amplitudes as
\begin{equation}
\label{eq:factorization1}
M_n|_{X_{ij}=0}=M_L\frac{1}{X_{ij}}M_R.
\end{equation} 
From the factorization property, the weights are constrained as
\begin{equation}
\sum_{P\in (ij)}\alpha_P=\sum_{P_L,P_R}\alpha_{P_L}\alpha_{P_R}
\end{equation} 
The left hand side involves sum over all accordiohedra $\mathcal{AC}_{n}^{P}$ for which $P\in (ij)$ and the right hand side involves sum over $P_L$ and $P_R$ which range over all the dissection of the two polygons $\{i,i+1,\dots,j\}$ and $\{j,j+1,\dots,n,1,\dots,i\}$ respectively. \par
We can further constrain the weights by repeating this procedure. By choosing the diagonal $(kl)\in\{i,i+1,\dots,j\}$, the polygon $\{i,i+1,\dots,j\}$ is divided into two polygons $\{k,k+1,\dots,l\}$ and $\{l,l+1,\dots,i,j,\dots,k\}$. We then obtain the following  
\begin{equation}
\label{eq:factorization2}
M_{i,i+1,\dots,j}|_{X_{kl}=0}=M_{L_2}\frac{1}{X_{kl}}M_{R_2}.
\end{equation} 
where $M_{L_2}$ and $M_{R_2}$ are sub-amplitudes $\{M_{L_2},M_{R_2}\}=\{M_{|l-k+1|}, M_{i+j+1-|l-k+1|}\}$. From \eqref{eq:factorization1} and \eqref{eq:factorization2}, we obtain that 
\begin{equation}
M_n|_{X_{ij}=0,X_{kl}=0}=M_{L2}\frac{1}{X_{kl}}M_{R_2}\frac{1}{X_{ij}}M_R.
\end{equation} 
Then the weights satisfy 
 \begin{equation}
\sum_{P\in (ij),(kl)}\alpha_P=\sum_{P_{L_2},P_{R_2},P_R}\alpha_{P_{L_2}}\alpha_{P_{R_2}}\alpha_{P_R}.
\end{equation} 
The left hand side involves sum over all accordiohedra $\mathcal{AC}_{n}^{P}$ for which $P\in (ij),(kl)$ and the right hand side involves sum over $P_{L_2}, P_{R_2}$ and $P_R$ which range over all the dissections of the three polygons $\{k,k+1,\dots,l\}$, $\{l,l+1,\dots,i,j,\dots,k\}$ and $\{j,j+1,\dots,n,1,\dots,i\}$ respectively. Continuing this procedure, we obtain 
\begin{equation}
M_n|_{X_{i_1j_1},X_{i_2j_2},\dots,X_{i_k,j_k}=0}=M_{L_k}\frac{1}{X_{i_kj_k}}M_{R_k}\frac{1}{X_{i_{k-1}j_{k-1}}}M_{R_{k-1}}\cdots M_{R_2}\frac{1}{X_{i_1j_1}}M_{R_1},
\end{equation} 
and
\begin{equation}
\label{eq:weightconstraints}
\sum_{P\in (i_1j_1),(i_2j_2),\dots,(i_kj_k)}\alpha_P=\sum_{P_{L_k},P_{R_k},\dots P_{R_1}}\alpha_{P_{L_k}}\alpha_{P_{R_k}}\cdots\alpha_{P_{R_1}}.
\end{equation} 
When the diagonals $(i_1j_1),(i_2j_2),\dots,(i_kj_k)$ make the complete dissection, this constraint takes the form :
\begin{equation}
\sum_{P\in (i_1j_1),(i_2j_2),\dots,(i_kj_k)}\alpha_P=1
\end{equation}
Since the complete dissection $(i_1j_1),(i_2j_2),\dots,(i_kj_k)$ is just the vertex of the accordiohedron, this constraint means that the residue in each channel $\prod_a\frac{1}{X_{i_a,j_a}}$ is one. This corresponds to the condition on weights derived from the relation between the weighted sum and the scattering amplitude in \cite{Alok,Raman} . However, here we derive this condition from the factorization property of the accordiohedron. For Stokes polytopes the weights have been derived using generalized BCFW recursion relations in \cite{Ishan}.
\subsection{Explicit calculations}
\subsubsection*{$n=5$ in $\phi^3+\phi^4$}
Here we determine the weights for $\phi^3+\phi^4$ case explicitly. The first nontrivial case is $n=5$. There is only way to divide the pentagon $\{1,2,\dots,5\}$ into one square and one triangle. Both of these have trivial weights, i.e.
\begin{equation}
\begin{split}
2\alpha_{5}&=1\\
\alpha_{5}&=\frac{1}{2}.
\end{split}
\end{equation}
\subsubsection*{$n=6$ in $\phi^3+\phi^4$}
\begin{figure}[H]
\centering
\includegraphics[width=180mm]{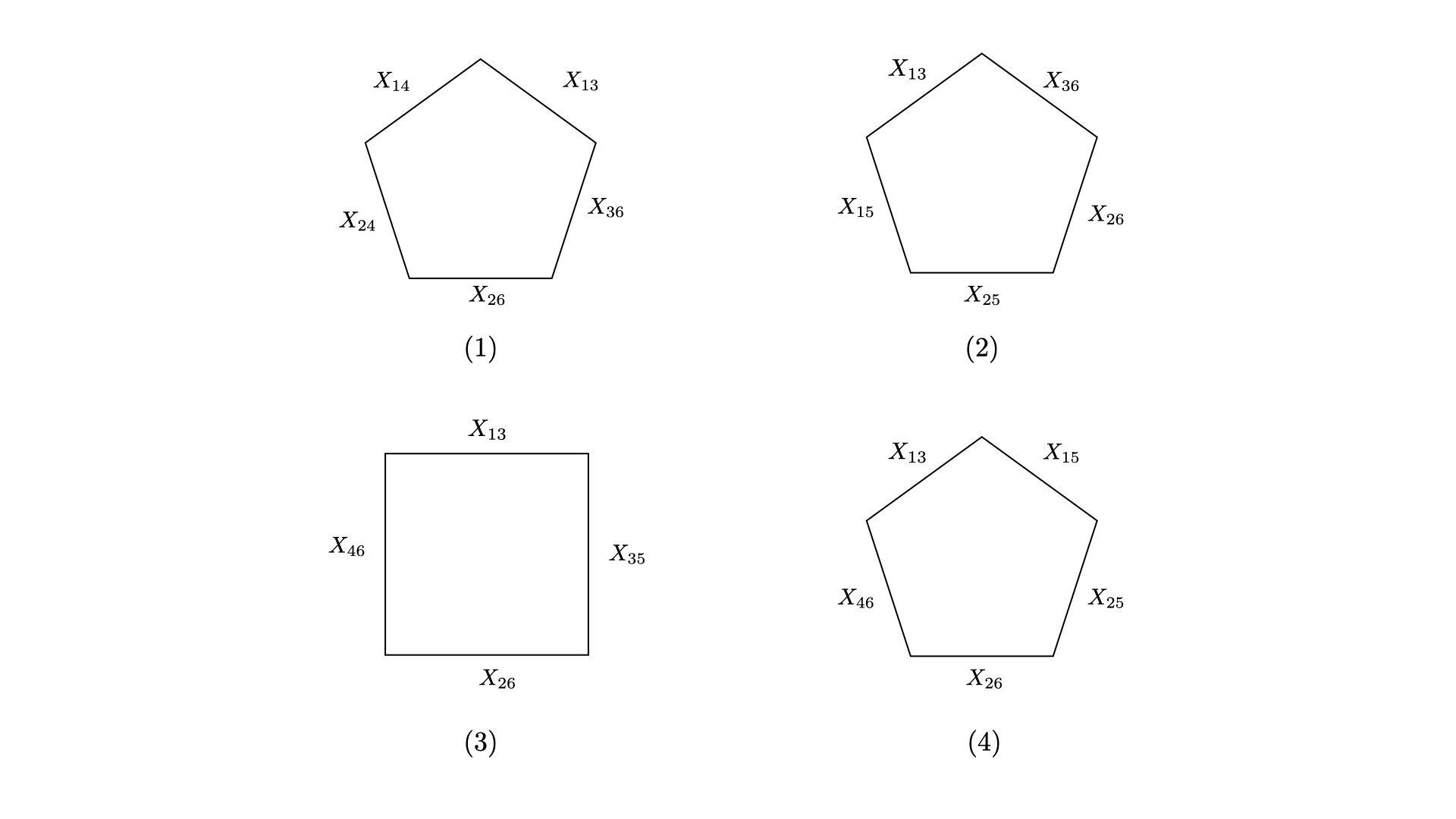}
\caption{Four accordiohedra for $n=6$.}
\label{fig:fouraccordiohedron}
\end{figure}

In this case there are four primitive accordiohedra, one square and three pentagons $\mathcal{AC}_{6}^{1},\dots,\mathcal{AC}_{6}^{4}$ as in Figure \ref{fig:fouraccordiohedron}. 
First we choose the diagonal $\{15\}$. Then the equation \eqref{eq:weightconstraints} gives a constraint on the weights as
\begin{align}
3\alpha_{1}+3\alpha_{2}+2\alpha_3+4\alpha_4=\frac{5}{2}.
\end{align}
 Similarly we can obtain the constraints from other diagonals. For example, we choose the diagonals $\{(13),(14)\}$, $\{(13),(46)\}$ and $\{(13),(15)\}$. The constraints are then given by :
\begin{align}
2\alpha_{1}+2\alpha_{2}+0\alpha_3+\alpha_4&=1\\
0\alpha_{1}+0\alpha_{2}+2\alpha_3+2\alpha_4&=1\\
\alpha_{1}+\alpha_{2}+\alpha_3+2\alpha_4&=1,
\end{align}
where $\alpha_1,\dots,\alpha_4$ are the weights for each accordiohedron respectively.
 From these constraints, we can obtain the weights 
 \begin{align}
\alpha_{1}=\frac{1}{2}-a,\ \alpha_{2}=a,\ \alpha_3=\frac{1}{2},\ \alpha_4=0.
\end{align}
This matches the result in \cite{Mrunmay}. If we choose other diagonals, the result does not change.
\subsubsection*{$n=7$ in $\phi^3+\phi^4$}
In this case there are $12$ topologically inequivalent Feynman diagrams  Figure \ref{fig:n7diagrams} and as many primitive accordiohedra. The reference dissections are :
\begin{figure}[H]
\centering
\includegraphics[width=180mm]{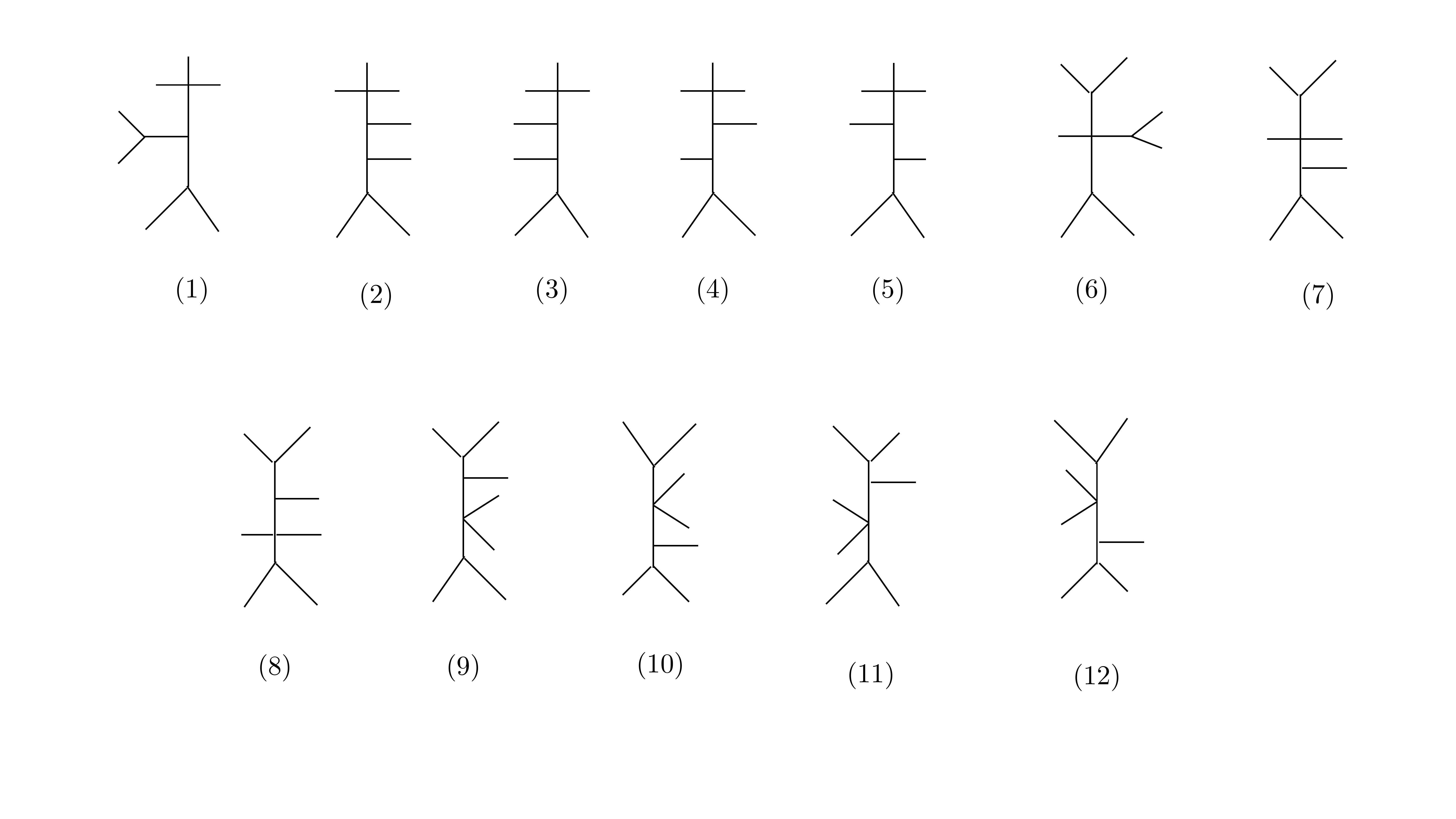}
\caption{12 topologically inequivalent Feynman diagrams.}
\label{fig:n7diagrams}
\end{figure}
\begin{align}
&1: \{14,16,46\}, 2: \{14,15,16\}, 3: \{14,46,47\}, 4: \{14,15,57\}, 5: \{14,47,57\}, 6: \{13,35,57\},\nonumber\\
 &7: \{13,47,57\}, 8: \{13,14,57\}, 9: \{13,14,16\}, 10: \{13,15,16\}, 11: \{13,14,46\},12: \{13,36,46\}.  
\end{align}
We determine the weights from factorization and obtain :
\begin{align}
\alpha_1+\alpha_2+\dots+\alpha_5=\frac{1}{2}, \alpha_7+\alpha_8=\frac{1}{2}, \alpha_6=\alpha_9=\alpha_{10}=\alpha_{11}=\alpha_{12}=0.
\end{align}
\subsubsection*{$n=8$ in $\phi^3+\phi^4+\phi^5$}
\begin{figure}[H]
\centering
\includegraphics[width=180mm]{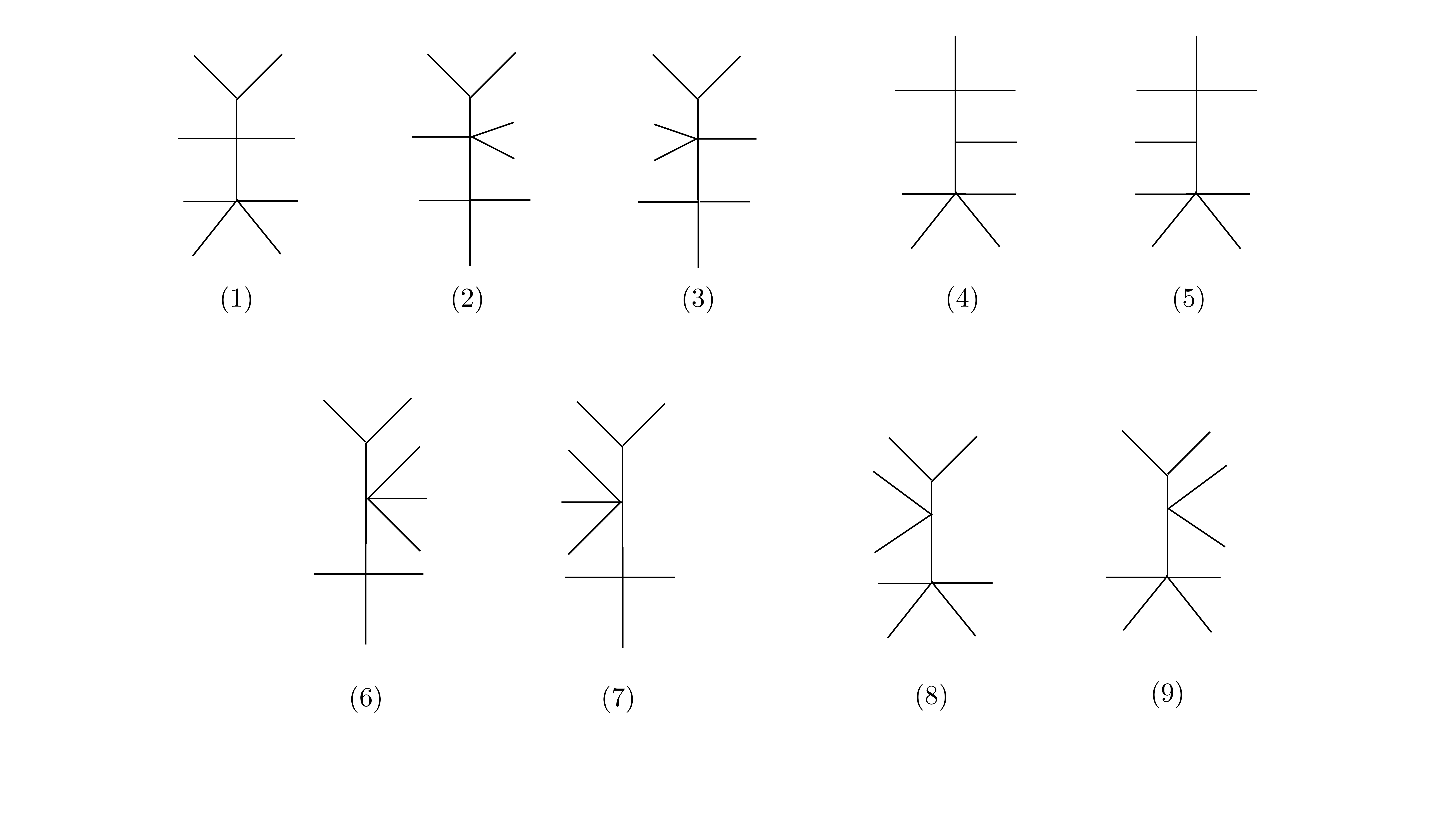}
\caption{Nine topologically inequivalent Feynman diagrams.}
\label{fig:n8diagrams}
\end{figure}

\begin{figure}[H]
\centering
\includegraphics[width=180mm]{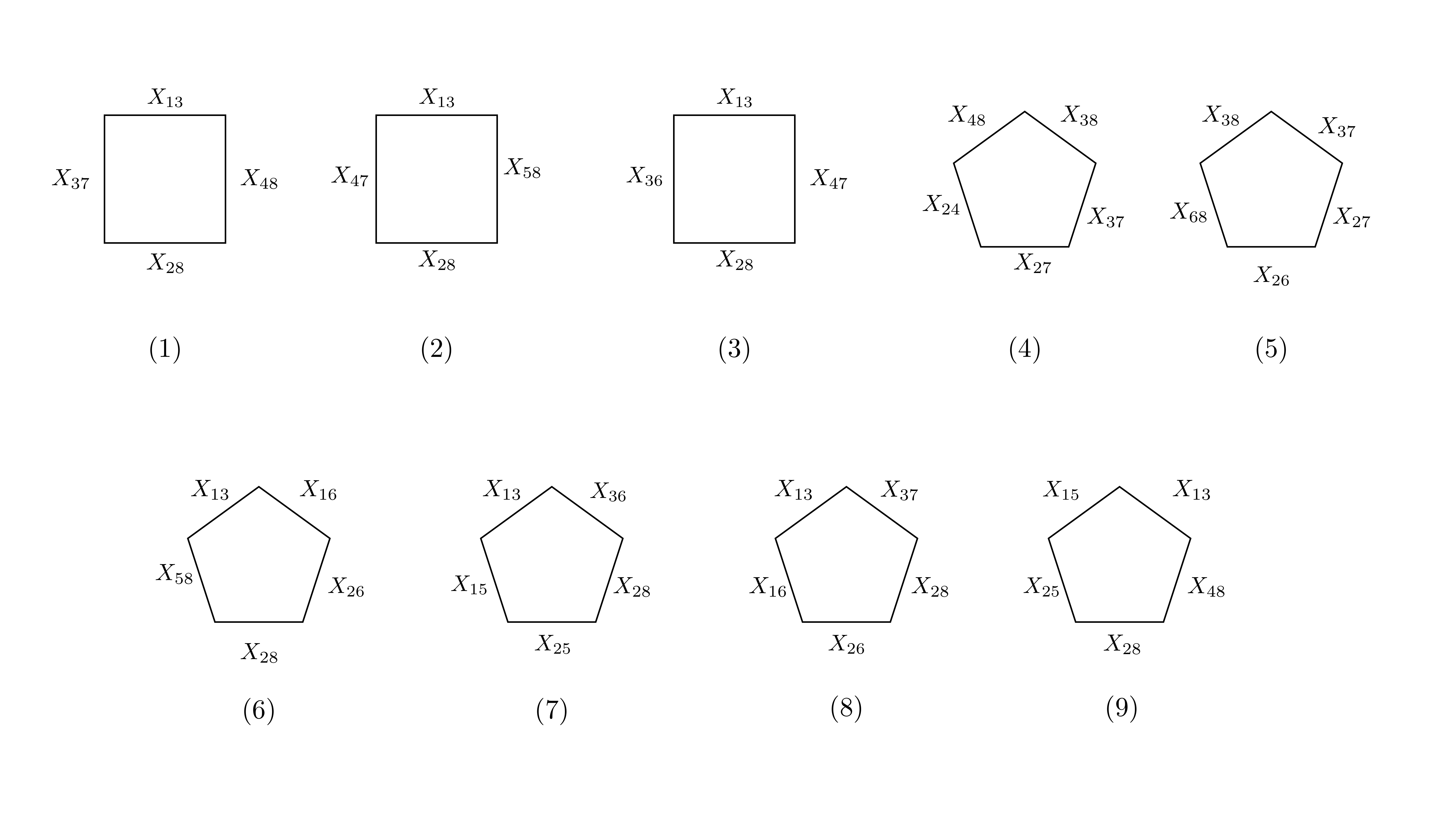}
\caption{Nine accordiohedra for $n=8$ case.}
\label{fig:345n8accordiohedron}
\end{figure}

Next, we consider the 8-point amplitude of $\phi^3+\phi^4+\phi^5$ theory. This amplitude is obtained from two type of interactions: two $\phi^4$ vertices or a $\phi^3$, a $\phi^4$, and a $\phi^5$ vertex each. Here we consider the latter. There are nine topologically inequivalent Feynman diagrams Figure \ref{fig:n8diagrams} and nine primitive accordiohedra Figure \ref{fig:345n8accordiohedron}.
We can easily obtain the canonical form for each accordiohedron. For example, the ratio part of the canonical form of the $(1)$ accordiohedron is 
\begin{align}
M^{(1)}=\frac{1}{X_{13}X_{48}}+\frac{1}{X_{48}X_{28}}+\frac{1}{X_{28}X_{37}}+\frac{1}{X_{37}X_{13}}.
\end{align}
To determine the weights, we choose the diagonals as $\{(13),(48)\}$, $\{(13),(58)\}$, $\{(13),(47)\}$, $\{(48),(83)\}$, $\{(83),(37)\}$, $\{(13),(16)\}$, $\{(13),(36)\}$, $\{(13),(37)\}$, $\{(13),(15)\}$. The constraints are given as 
\begin{align}
2\alpha_{1}+\alpha_{8}+\alpha_9&=1\nonumber\\
2\alpha_{2}+\alpha_{3}+\alpha_6&=1\nonumber\\
\alpha_{2}+2\alpha_{3}+\alpha_7&=1\nonumber\\
2\alpha_{4}+\alpha_{5}+\alpha_6+\alpha_8&=1\nonumber\\
\alpha_{4}+2\alpha_{5}+\alpha_7+\alpha_9&=1\\
\alpha_{2}+\alpha_{4}+2\alpha_6+\alpha_8&=1\nonumber\\
\alpha_{3}+\alpha_{5}+2\alpha_7+\alpha_9&=1\nonumber\\
\alpha_{1}+\alpha_{4}+\alpha_6+2\alpha_8&=1\nonumber\\
\alpha_{1}+\alpha_{5}+\alpha_7+2\alpha_9&=1\nonumber.
\end{align}
where $\alpha_1,\dots,\alpha_9$ are the weights for each accordiohedron respectively. From these constraints, we can obtain the weights 
 \begin{align}
\alpha_{1}=\frac{1}{2},\ \alpha_{2}=a,\ \alpha_3=\frac{1}{2}-a,\ \alpha_4=a,\ \alpha_5=\frac{1}{2}-a,\ \alpha_6=\frac{1}{2}-a,\ \alpha_7=a,\ \alpha_8=0,\ \alpha_9=0.
\end{align}
The weighted sum of the ratio parts of the canonical forms with these weights corresponds to the $n=8$ amplitude.
\section{Recursion relation for $\phi^p$ amplitudes}
\label{derivation}
In this section we review recursion relations for planar tree level amplitudes in scalar theories with $\phi^p$ interaction where $p\ge 4$ \cite{Ryota}. This is a direct generalisation of the derivation of recursion relations for $\phi^3$ theories that was obtained in \cite{Song-Yang,Yang}. Further we claim in this paper the recursions are valid for theories with polynomial interactions.

A $d$-dimensional polytope $A$ in kinematic space is described by as many independent planar kinematic variables $X_{A_{i}}$ where $ i=1,\ldots,d$. These form the basis variables in terms of which the remaining planar kinematic variables are expressed. 
Let us now consider the following rescaling of a subset of basis variables :
\begin{align}
\label{shift}
X_{A_i} \to z\,X_{A_i},\quad\quad i=1,2,\ldots,k\quad\text{and}\quad k\le d
\end{align}
%
Under this rescaling of basis variables, a subset of the dependent variables are also deformed. We denote the deformed $n$-point amplitude as $A_n(zX,C)$.
Let us now consider the following integral :
\begin{align}
\label{integrand}
\oint dz \frac{z^k}{z-1} A_n(zX,C)
\end{align}
By construction the original undeformed amplitude is given by the residue of the integrand at $z=1$. Cauchy's residue theorem expresses the contour integral as the sum of residues from the pole at infinity and finite poles :
\begin{align}
\label{Cauchy}
A_n(X,C)=-\left(\text{Res}_{\infty} +\sum_{\text{finite\,poles}} \text{Res}_{z_i}\right)\,\left(\frac{z^k}{z-1} A_n(zX,C)\right)
\end{align}
We will soon show that the integrand in \eqref{integrand} has no pole at $z=0$ or $z=\infty$.
%
%

A canonical convex realisation of the associahedron in the kinematic space for $\phi^3$ theories was given in \cite{ABHY}. A rigorous derivation of the equations that give a convex realisations of accordiohedra was obtained in \cite{1906ppp} (see also \cite{Ours}). In short, these equations take the form :
\begin{align}
\label{compatibleX}
X_{B_i}= \epsilon_{B_i}+X_i+ \sum_{j=1}^k \lambda_{ij} X_{A_j}
\end{align}
where $X_{B_i}$ denote planar variables compatible with the basis variables, $\epsilon_{B_i}$ and $X_i$ denote respectively suitable linear combinations of constants that appear in the convex realisation (the detailed knowledge of which is not required in this work) and of the undeformed basis variables $\{X_{A_j}\}_{ j=k+1,...,d}$ respectively and $\lambda_{ij}$ are real constants.


The poles of the deformed amplitude $A_n(zX,C)$ is given by solutions of equations $\hat X_{B_i}(z)=0$ where $\hat X_{B_i}(z)$ denotes the deformed $X_{B_i}$ under the shift \eqref{shift}.\\
%
At physical poles, amplitude factorizes as :
\begin{equation}
\label{factorization}
\text{limit}_{\hat X_{B_i} \to 0}~ \hat X_{B_i}(z) A_n(zX,C) = A_{B_i}^{\text{limit}}(z_{B_i}X,C)
\end{equation}
$A_{B_i}^{\text{limit}}$ is given by the product of lower point amplitudes.
For example, in the case of tree amplitudes \eqref{factorization} takes the form:

\begin{equation}
\text{limit}_{X_{pq} \to 0} \, X_{pq} \,A_{1,2,...,n}(X,C)= A_{p,...,q-1,I} \times A_{I,q,...,p-1}
\end{equation}
with $I$, the internal particle going on-shell.
The residue then reads,
\begin{equation}
\text{Res}_{\hat X_{B_i}} \frac{z^k}{z-1}A_n(zX,C)=\frac{z_i^k}{(\sum_{j=1}^k\lambda_{ij}X_{A_j})(z_i-1)}A_{B_i}^{\text{limit}}(z_{B_i}X,C)
\end{equation}
This leads to the recursion formula for $n$-point amplitude with a given quadragulation (or an appropriate dissection in general)
\begin{align}
\label{generalscalingformula}
A_n(X,C)=\sum_{B_i}\frac{z_i^k}{X_{B_i}}A_{B_i}^{\text{limit}}(z_{B_i}X,C)
 \end{align}
For the case of one variable rescaling $X_A \to z X_A$, supposing $X_{B_i} \propto \lambda_i X_A$ the above equation takes the following simple form  
\begin{align}
 \label{singlescalingformula}
A_n(X,C)=\sum_{B_i} \left(\frac{1}{X_{B_i}}-\frac{1}{\lambda_i X_A}\right)  A_{B_i}^{\text{limit}}(z_{B_i}X,C)
 \end{align}
Further for this case  evaluating function $A_{B_i}^{\text{limit}}(z_{B_i}X,C)$ at each $z_{B_i}$  translates to doing the following substitution in $A_{B_i}^{\text{limit}}(X,C)$ :
$X_p\rightarrow X_p − \frac{1}{\lambda_i}\,X_{B_i}$ for each $X_{B_i}\propto \lambda_{i}\,X_{p}$, where $X_p$ are the deformed kinematic variables (including the basis variables).

\subsection{Proof for no pole at $z=0$}
The canonical function of simple polytopes is given by \cite{Salvatori:2019phs},
\begin{equation}
\label{canonical_function}
\underline{\Omega_p} = \sum_{v\,\in\,\text{vertices}}\,\,\prod_{\text{facets}\,f\,\in\,v} \frac{1}{X_f}
\end{equation}
For a $d$ dimensional simple polytope, there are exactly $d$ facets adjacent to any vertex $v$. Thus, the product $\prod \frac{1}{X_f}$ is proportional to $\frac{1}{X^d}$. If we rescale $k$ of the basis variables, among the $d$ $X_{ij}$'s that appear in the product in \eqref{canonical_function} atmost $k$ of them have a $z$ dependence that leads to a pole at $z=0$. Non-basis variables do not give rise to a pole at $z=0$. Thus the product in \eqref{canonical_function} can at most give rise to an order $k$ pole at $z=0$ and that is precisely cancelled by the $z^k$ factor in the integrand \eqref{integrand}. This proves that the integrand in \eqref{integrand} has no pole at $z=0$.

\subsection{Proof for no pole at infinity}
%
%
%

To obtain the $O(\frac{1}{z^k})$ contribution, we need $k$ of the $X_{ij}$ in the denominator of \eqref{canonical_function} shifted and $(d-k)$ of them undeformed. If we denote the undeformed ones as $\widetilde X_j$, we have the following contribution to the amplitude :
\begin{equation}
\left(\sum_{\text{vertices shared by}\,(d-k)\,\widetilde X_j} \frac{1}{\prod_{k} X}\right) \frac{1}{\prod_{j=1}^{d-k} \widetilde X_j}
\end{equation}
The quantity in the bracket is the canonical function for the ABHY polytope when all the $\widetilde X_j \to 0$ and we denote it by $A_k(X,C)$. Since all the $X$'s in $A_k$ are shifted, as $z \to \infty$  the leading contribution is given by setting $C \to 0$ :
\begin{equation}
\text{limit}_{z \to \infty} A_k(zX,C) \sim \frac{1}{z^k} A_k(X,0)=0
\end{equation} 
This fact (known as the soft limit) can be easily seen, as $A_k(X,0)$ is always given by product of lower point functions which can be explicitly checked to be zero e.g. $n=6$ quartic amplitude with $Q=(14)$ is given by $A_6=(\frac{1}{X_{14}}+\frac{1}{-X_{14}+C})$ which vanishes as $C \to 0$. Geometrically it implies that the size of the accordiohedron has shrunk to zero.

Thus we see that the integrand in \eqref{integrand} has no pole at infinity. This completes our derivation of the recursion relation.

Note that although this derivation of recursion relation is identical to the one given in \cite{Yang,Ryota},  we have made use of the canonical convex realization \eqref{compatibleX} of accordiohedra in kinematic space. This makes the derivation of the recursion relation independent of any geometric picture of associahedron.


\section{Explicit computations in $\phi^p$ theories}
\label{Examples}
We will now see  explicit examples where we show that the recursion relations \eqref{generalscalingformula} reproduce the correct amplitude. 
The power of recursion relations in $\phi^p$ theories using specific embeddings of a few accordiohedra was studied in \cite{Ryota}. To further illustrate the power of the recursion relations, we use them for a wide class of theories with mixed vertices. In our analysis we make use of the canonical convex embedding of the relevant accordiohedron in the positive region of the kinematic space \cite{1906ppp,Ours}.

\subsection{$n=10$ in $\phi^4$}
\label{n=10 amplitude}
We will start by discussing  the $n=10$ case in $\phi^4$  theory. It has been shown that there are four families of three-dimensional Stokes polytopes based on the reference quadrangulation of a decagon with cyclically ordered vertices \cite{Alok}. These are the cube type, the snake type, the Lucas type and the mixed type.
While the cube type and the snake type are the three dimensional generalisation of the square and the pentagon which appear in the $n=8$ case, the mixed type and Lucas type are new polytopes that appear in $n=10$. 

In the following we will consider the recursion relations for each of these cases when two of the three basis variables are rescaled. While interesting in its own right, it will also serve us with enough data to obtain a better understanding of projective triangulations which we will come to in Section~\ref{triangulations}.

\subsubsection{Cube}
\label{cube}
The cube type Stokes polytope arises from the reference quadrangulation $Q=\{14,(5,10),69\}$ of a decagon with cylically ordered vertices from 1 to 10. The $Q$-compatible diagonals are $\{49,58,(3,10)\}$. The embedding of the polytope in the positive kinematic space is given by the following constraints between the planar variables :
\begin{align}
\label{n=10cuberelations}
X_{49}&=-X_{(5,10)}+\epsilon_{49}\cr
X_{58}&=-X_{69}+\epsilon_{58}\cr
X_{(3,10)}&=-X_{14}+\epsilon_{(3,10)}
\end{align}
We will now consider the scaling of two of the three basis variables. We choose them to be $X_{14}$ and $X_{(5,10)}$ :
\begin{align}
X_{14}\rightarrow z\,X_{14},\quad\quad X_{(5,10)}\rightarrow z\,X_{(5,10)}\,.
\end{align}
From \eqref{n=10cuberelations}, we see that under this scaling two of the dependent variables namely $X_{(3,10)}$ and $X_{49}$ are deformed. The recursion formula \eqref{generalscalingformula} then takes the form :
\begin{align}
\label{cubeeqn}
M_{10}^{\{14,(5,10),69\}}=\frac{\hat z_{(3,10)}^2} {X_{(3,10)}}& \left(\frac{1}{ \hat z_{(3,10)} X_{(5,10)} ~X_{69}} + \frac{1}{ \hat z_{(3,10)} X_{(5,10)} ~X_{58}}+ \frac{1}{ \widehat X_{49,(3,10)} ~X_{69}} + \frac{1}{ \widehat X_{49,(3,10)} ~X_{58}} \right)\nonumber\\[5pt]
&+ \frac{\hat z_{49}^2} {X_{49}} \left(\frac{1}{ \hat z_{49} X_{14} ~X_{69}} + \frac{1}{ \hat z_{49} X_{14} ~X_{58}}+ \frac{1}{ \widehat X_{(3,10),49} ~X_{69}} + \frac{1}{ \widehat X_{(3,10),49} ~X_{58}} \right)
\end{align}
where
\begin{align}
\hat z_{(3,10)}=&-\frac{\epsilon_{(3,10)}}{X_{14}},\quad  \hat z_{49}=\frac{\epsilon_{49}}{X_{(5,10)}},\nonumber\\[5pt]
\widehat {X}_{49,(3,10)}&=\epsilon_{49}-\hat z_{(3,10)}\,X_{(5,10)}
\end{align}
One can now easily check that $M_{10}^{\{14,(5,10),69\}}$ matches the expected result for the partial amplitude corresponding to the reference quadrangulation $Q=\{14,(5,10),69\}$\cite{Alok}.
%
%
\subsubsection{Snake}
\label{snake}
The snake type Stokes polytope corresponds to the reference quadrangulation $Q=\{14,16,18\}$ of the decagon. The $Q$-compatible diagonals are $\{36,58,(7,10),38,(5,10),(3,10)\}$. The embedding of the polytope in the positive kinematic space is given by the constraints :
\begin{align}
\label{snake_constraints}
X_{36}&=\epsilon_{36}-X_{14}+X_{16}\cr
X_{58}&=\epsilon_{58}-X_{16}+X_{18}\cr
X_{(7,10)}&=\epsilon_{(7,10)}-X_{18}\cr
X_{38}&=\epsilon_{38}-X_{14}+X_{18}\cr
X_{(5,10)}&=\epsilon_{(5,10)}-X_{16}\cr
X_{(3,10)}&=\epsilon_{(3,10)}-X_{14}
\end{align}
Let us now consider the scaling of two basis variables, say $X_{14}$ and $X_{16}$ :
\begin{align}
X_{14}\rightarrow z\,X_{14},\quad X_{16}\rightarrow z\,X_{16}\,.
\end{align}
From \eqref{snake_constraints} we see that under this scaling, the dependent  planar variables  $X_{36},X_{58},X_{38},X_{(5,10)}$ and $X_{(3,10)}$ are deformed. The recursion formula \eqref{generalscalingformula} takes the form :
\begin{tiny}
\begin{align}
\label{snakeeqn}
M_{10}^{\{14,16,18\}}&=\frac{\hat z_{36}^2} {X_{36}} \left(\frac{1}{ \hat z_{36} X_{16} ~X_{18}} + \frac{1}{ \hat z_{36} X_{16} ~X_{(7,10)}}+ \frac{1}{ \widehat X_{38,36} ~X_{18}} + \frac{1}{ \widehat X_{(3,10),36} ~X_{(7,10)}}+ \frac{1}{ \widehat X_{38,36} ~\widehat X_{(3,10),36}} \right) \cr
&\hspace{.2cm}+ \frac{\hat z_{58}^2} {X_{58}}\left(\frac{1}{ \hat z_{58} X_{14} ~X_{18}} + \frac{1}{ \hat z_{58} X_{14} ~\widehat X_{(5,10),58}}+ \frac{1}{ \widehat X_{38,58} ~X_{18}} + \frac{1}{ \widehat X_{(3,10),58} ~\widehat X_{(5,10),58}}+ \frac{1}{ \widehat X_{38,58} ~\widehat X_{(3,10),58}} \right)\cr
&\hspace{.2cm}+ \frac{\hat z_{38}^2} {X_{38}}\left(\frac{1}{ \widehat X_{36,38} ~X_{18}} + \frac{1}{ \widehat X_{58,38} ~X_{18}}+ \frac{1}{ \widehat X_{36,38} ~\widehat X_{(3,10),38}} + \frac{1}{ \widehat X_{58,38} ~\widehat X_{(3,10),38}} \right)\cr
&\hspace{.5cm}+ \frac{\hat z_{(5,10)}^2} {X_{(5,10)}}\left(\frac{1}{ \hat z_{(5,10)} X_{14} ~\widehat X_{58,(5,10)}} + \frac{1}{ \hat z_{(5,10)} X_{14} ~X_{(7,10)}}+ \frac{1}{ \widehat X_{(3,10),(5,10)} ~\widehat X_{58,(5,10)}} + \frac{1}{ \widehat X_{(3,10),(5,10)} ~ X_{(7,10)}} \right)\cr
&\hspace{.2cm}+ \frac{\hat z_{(3,10)}^2} {X_{(3,10)}}\left(\frac{1}{ \widehat X_{36,(3,10)} ~X_{(7,10)}} + \frac{1}{ \widehat X_{36,(3,10)} ~\widehat X_{38,(3,10)}}+ \frac{1}{ \widehat X_{58,(3,10)} ~\widehat X_{(5,10),(3,10)}} + \frac{1}{ \widehat X_{38,(3,10)} ~\widehat X_{58,(3,10)}}+ \frac{1}{\widehat X_{(5,10),(3,10)} ~X_{(7,10)}} \right)\cr
\end{align}
\end{tiny}
where
\begin{align}
\hat z_{36}=\frac{\epsilon_{36}}{X_{14}-X_{16}},\quad\hat z_{58}=\frac{\epsilon_{58}+X_{18}}{X_{16}},\quad\hat z_{38}=\frac{\epsilon_{38}+X_{18}}{X_{14}},\quad\hat z_{(5,10)}=\frac{\epsilon_{(5,10)}}{X_{16}},\quad \hat z_{(3,10)}=\frac{\epsilon_{(3,10)}}{X_{14}}
\end{align}
and $\widehat X_{ij,kl}$ stands for the dependent variable $X_{ij}$ deformed by $\hat z_{kl}$. It is straightforward to check that $M_{10}^{\{14,16,18\}}$ matches the expected result for the partial amplitude corresponding to reference quadrangulation $Q=\{14,16,18\}$ \cite{Alok}.
%
%
\subsubsection{Lucas type}
\label{lucas}
The Lucas type Stokes polytope corresponds to the reference quadrangulation $Q=\{14,47,(7,10)\}$ of the decagon. The $Q$-compatible diagonals are $\{16,(3,10),36,49,69\}$.
The embedding of the Lucas type polytope in the positive region of kinematic space is given by the constraints :
\begin{align}
\label{lucas_constraints}
X_{36}&=-X_{47}+\epsilon_{36}\cr
X_{69}&=-X_{(7,10)}+\epsilon_{69}\cr
X_{(3,10)}&=-X_{14}+\epsilon_{(3,10)}\cr
X_{16}&=X_{14}-X_{47}+\epsilon_{16}\cr
X_{49}&=X_{47}-X_{(7,10)}+\epsilon_{49}\,.
\end{align}
Consider the scaling of the basis variables  $X_{14}$ and $X_{47}$ :
\begin{align}
X_{14}\rightarrow z\,X_{14},\quad X_{47}\rightarrow z\,X_{47}\,.
\end{align}
Under this scaling, we see from \eqref{lucas_constraints} that the dependent variables  $X_{16},X_{(3,10)},X_{36}$ and $X_{49}$ are deformed. The recursion formula \eqref{generalscalingformula} takes the form :
\begin{tiny}
\begin{align}
\label{lucaseqn}
M_{10}^{\{14,47,(7,10)\}}&=\frac{\hat z_{16}^2} {X_{16}}\left(\frac{1}{ \hat z_{16}X_{14} ~X_{(7,10)}} + \frac{1}{ \widehat X_{36,16} ~X_{(7,10)}}+ \frac{1}{ \hat z_{16}X_{14} ~\widehat X_{69,16}} + \frac{1}{ \widehat X_{36,16} ~\widehat X_{69,16}} \right)\cr
&\hspace{.1cm}+ \frac{\hat z_{(3,10)}^2} {X_{(3,10)}}\left(\frac{1}{ \hat z_{(3,10)} X_{47} ~ X_{(7,10)}} + \frac{1}{ \widehat X_{49,(3,10)} ~\hat z_{(3,10)} X_{47}}+ \frac{1}{ \widehat X_{36,(3,10)} ~ X_{(7,10)}} + \frac{1}{ \widehat X_{49,(3,10)} ~ \widehat X_{69,(3,10)}}+\frac{1}{\widehat X_{36,(3,10)} \widehat X_{69,(3,10)}} \right)\cr
&\hspace{.1cm}+\frac{\hat z_{36}^2} {X_{36}} \left(\frac{1}{ \widehat X_{(3,10),36} ~X_{(7,10)}} + \frac{1}{ \widehat X_{16,36} ~X_{(7,10)}}+ \frac{1}{ \widehat X_{(3,10),36} ~\widehat X_{69,36}} + \frac{1}{ \widehat X_{16,36} ~\widehat X_{69,36}} \right) \cr
&\hspace{.1cm}+ \frac{\hat z_{49}^2} {X_{49}}\left(\frac{1}{ \hat z_{49} X_{14} ~\hat z_{49}X_{47}} + \frac{1}{ \widehat X_{(3,10),49} ~\hat z_{49}X_{47}}+ \frac{1}{ \hat z_{49} X_{14}~\widehat X_{69,49}} + \frac{1}{ \widehat X_{(3,10),49} ~\widehat X_{69,49}} \right)
\end{align}
\end{tiny}
where
\begin{align}
\hat z_{16}=-\frac{\epsilon_{16}}{X_{14}-X_{47}},\quad\hat z_{(3,10)}=\frac{\epsilon_{(3,10)}}{X_{14}},\quad\hat z_{36}=\frac{\epsilon_{36}}{X_{47}},\quad\hat z_{49}=\frac{-\epsilon_{49}+X_{(7,10)}}{X_{47}}
\end{align}
and $\widehat X_{ij,kl}$ stands for the dependent variable $X_{ij}$ deformed by $\hat z_{kl}$. It can be easily checked that this matches the expected result for the partial amplitude corresponding to the reference quadrangulation $Q=\{14,47,(7,10)\}$ \cite{Alok}.
%
%
%
\subsubsection{Mixed type}
\label{mixed}
The mixed type Stokes polytope corresponds to the reference quadrangulation $Q=\{(14),(5,10),(7,10)\}$ of the decagon. The   $Q$-compatible diagonals are $\{49,69,(3,10),47\}$. The embedding of the polytope in the positive kinematic space is given by the following constraints :
\begin{align}
\label{n=10mixedrelations}
X_{49}&=-X_{(5,10)}+\epsilon_{49}\cr
X_{69}&=-X_{(7,10)}+\epsilon_{69}\cr
X_{(3,10)}&=-X_{14}+\epsilon_{(3,10)}\cr
X_{47}&=X_{(7,10)}-X_{(5,10)}+\epsilon_{47}\,.
\end{align}
Let us scale two basis variables $X_{14}$ and $X_{(5,10)}$  : 
\begin{align}
X_{14}\rightarrow z\,X_{14},\quad X_{(5,10)}\rightarrow z\,X_{(5,10)}\,.
\end{align}
Under this scaling, we see from \eqref{n=10mixedrelations} that the dependent variables  $X_{47},X_{(3,10)},X_{49}$ and $X_{49}$ are deformed. The recursion formula then \eqref{generalscalingformula} takes the form :
\begin{tiny}
\begin{align}
\label{mixedeqn}
M_{10}^{\{14,(5,10),(7,10)\}}&=\frac{\hat z_{(3,10)}^2} {X_{(3,10)}}\left(\frac{1}{ \hat z_{(3,10)}X_{(5,10)} ~X_{(7,10)}} + \frac{1}{ \widehat X_{47,(3,10)} ~X_{(7,10)}}+ \frac{1}{ \hat z_{(3,10)}X_{(5,10)} ~\widehat X_{69,(3,10)}} + \frac{1}{ \widehat X_{49,(3,10)} ~\widehat X_{69,(3,10)}}+ \frac{1}{ \widehat X_{47,(3,10)} ~\widehat X_{49,(3,10)}} \right)\cr
&+ \frac{\hat z_{47}^2} {X_{47}}\left(\frac{1}{ \hat z_{47} X_{14} ~ X_{(7,10)}} + \frac{1}{ \widehat X_{(3,10),47} ~ X_{(7,10)}}+ \frac{1}{ \hat z_{47} X_{14} ~ \widehat X_{49,47}} + \frac{1}{ \widehat X_{(3,10),47} ~ \widehat X_{49,47}} \right)\cr
&+\frac{\hat z_{49}^2} {X_{49}} \left(\frac{1}{ \hat z_{49} X_{14} ~\widehat X_{47,49}} + \frac{1}{ \widehat z_{49} X_{14} ~\widehat X_{69,49}}+ \frac{1}{ \widehat X_{(3,10),49} ~\widehat X_{69,49}} + \frac{1}{ \widehat X_{(3,10),49} ~\widehat X_{47,49}} \right) 
\end{align}
\end{tiny}
where
\begin{align}
\hat z_{(3,10)}=-\frac{\epsilon_{(3,10)}}{X_{14}},\quad\hat z_{47}=\frac{\epsilon_{47}+X_{(7,10)}}{X_{(5,10)}},\quad\hat z_{49}=\frac{\epsilon_{49}}{X_{(5,10)}}
\end{align}
and $\widehat X_{ij,kl}$ stands for the dependent variable $X_{ij}$ deformed by $\hat z_{kl}$. This matches the result obtained for the partial amplitude corresponding to the reference quadrangulation $Q=\{14,(5,10),(7,10)\}$ in \cite{Alok}.
%
%

\subsection{Mixed vertices}
\label{poly_int}
In this section, we illustrate the validity of the recursion relation \eqref{generalscalingformula} when we have a theory with mixed vertices. The positive geometry associated to these theories have been studied in \cite{Mrunmay}. As in the case of the Stokes polytopes in the $\phi^4$ theory we will make use of the canonical convex embedding of the relevant accordiohedron in each case \cite{1906ppp,Ours}.
\subsection*{$n=6$ in $\phi^3+\phi^4$}
In this case there are four accordiohedra as shown in Figure \ref{fig:fouraccordiohedron}. We consider the first case with reference dissection $D=\{13,14\}$ and the corresponding accordiohedron $\mathcal{AC}_1^6$ is a pentagon. The compatible diagonals are $\{36,26,24\}$. The embedding of the polytope in the positive kinematic space is given by the constraints :
\begin{align}
\label{n=6mixedrelations}
X_{36}&=-X_{14}+\epsilon_{36}\cr
X_{26}&=-X_{13}+\epsilon_{26}\cr
X_{24}&=X_{14}-X_{13}+\epsilon_{24}
\end{align}
We consider the scaling of a single basis variable :
\begin{equation}
X_{13} \to z\,X_{13}
\end{equation}
Under this scaling the set of deformed variables from \eqref{n=6mixedrelations} is $\{X_{24},X_{26}\}$. The recursion formula for single variable scaling \eqref{singlescalingformula} then takes the form:
\begin{align}
M_6^{(13,14)}=\left( \frac{1}{X_{24}}+\frac{1}{X_{13}}\right) \left(\frac{1}{X_{14}}+\frac{1}{X_{26}-X_{24}}\right)+\left(\frac{1}{X_{26}}+\frac{1}{X_{13}}\right) \left(\frac{1}{X_{24}-X_{26}}+\frac{1}{X_{36}}\right)
\end{align}
This matches the expected partial amplitude corresponding to the reference dissection $D=\{13,14\}$ :
\begin{align}
\frac{1}{X_{24}X_{14}}+\frac{1}{X_{13}X_{14}}+\frac{1}{X_{24}X_{26}}+\frac{1}{X_{26}X_{36}}+\frac{1}{X_{13}X_{36}} 
\end{align}
It is easy to show that for other three accordiohedra in this case, the recursion relation reproduces the correct partial amplitude.
\subsection*{$n=7$ in $\phi^3+\phi^4$}
In this case there are 12 accordiohedra. We consider the first case with reference dissection $D=\{14,16,46\}$. The diagonals compatible with this dissection are $\{15,35,36,37,47,57\}$. The embedding of the polytope in the positive kinematic space is given by the constraints :
\begin{align}
\label{n=7mixedrelations}
X_{15}&=X_{14}-X_{46}+\epsilon_{15}\cr
X_{35}&=-X_{46}+\epsilon_{35}\cr
X_{36}&=X_{16}-X_{14}+\epsilon_{36}\cr
X_{37}&=-X_{14}+\epsilon_{37}\cr
X_{47}&=X_{46}-X_{16}+\epsilon_{47}\cr
X_{57}&=-X_{16}+\epsilon_{57}
\end{align}
Let us now consider the scaling of two of the three basis variables :
\begin{equation}
X_{14} \to z\,X_{14}, \quad X_{16} \to z\,X_{16}
\end{equation}
The set of deformed variables under this scaling from \eqref{n=7mixedrelations} is $\{X_{57},X_{15},X_{36},X_{37},X_{47}\}$. The recursion formula \eqref{generalscalingformula} takes the form :
\begin{align}
\label{n=7eqn}
A_7^{(14,16,46)}=&\frac{\hat z_{57}^2} {X_{57}} \left(\frac{1}{ \hat z_{57} X_{14} ~\widehat X_{47,57}} + \frac{1}{\widehat X_{37,57}  ~\widehat X_{47,57}}+ \frac{1}{ \hat z_{57} X_{14} ~\widehat X_{15,57}} + \frac{1}{ \widehat X_{15,57} ~X_{35}}+ \frac{1}{  X_{35} ~\widehat X_{37,57}} \right) \cr
&+ \frac{\hat z_{15}^2} {X_{15}}\left(\frac{1}{ \hat z_{15} X_{14} ~\hat z_{15} X_{16}} + \frac{1}{ \hat z_{15} X_{14} ~\hat X_{57,15}}+ \frac{1}{ \hat z_{15} X_{16} ~X_{35}} + \frac{1}{ \widehat X_{57,15} ~ X_{35}} \right)\cr
&+ \frac{\hat z_{36}^2} {X_{36}}\left(\frac{1}{ \hat z_{36} X_{16} ~X_{46}} + \frac{1}{ \hat z_{36} X_{16} ~X_{35}}+ \frac{1}{ \widehat X_{37,36} ~ X_{46}} + \frac{1}{ \widehat X_{37,36} ~X_{35}} \right)\cr
&+ \frac{\hat z_{37}^2} {X_{37}}\left(\frac{1}{ \widehat X_{36,37} ~ X_{46}} + \frac{1}{ \widehat X_{47,37} ~X_{46}}+ \frac{1}{ \widehat X_{36,37} ~ X_{35}} + \frac{1}{ \widehat X_{47,37} ~ X_{57,37}} +\frac{1}{X_{35} \widehat X_{57,37}}\right)\cr
&+ \frac{\hat z_{47}^2} {X_{47}}\left(\frac{1}{ \hat z_{47} X_{14} ~X_{46}} + \frac{1}{ \widehat X_{37,47} ~X_{46}}+ \frac{1}{ \hat z_{47} X_{14} ~\widehat X_{57,47}} + \frac{1}{ \widehat X_{37,47} ~\widehat X_{57,47}} \right)
\end{align}
where
\begin{align}
\hat z_{57}=\frac{\epsilon_{57}}{X_{16}},\quad\hat z_{15}=\frac{X_{46}-\epsilon_{15}}{X_{14}},\quad\hat z_{36}=\frac{\epsilon_{36}}{X_{14}-X_{16}},\quad\hat z_{37}=\frac{\epsilon_{37}}{X_{14}},\quad \hat z_{47}=\frac{\epsilon_{47}+X_{46}}{X_{16}}
\end{align}
and $\hat X_{ij,kl}$ as before stands for the dependent variable $X_{ij}$ deformed by $\hat z_{kl}$.
It can be easily checked that this matches the expected result for the partial amplitude corresponding to the principle dissection $D=\{14,16,46\}$ :
\begin{equation}
\begin{split}
\Bigg(&\frac{1}{X_{14}X_{16}X_{46}}+\frac{1}{X_{36}X_{16}X_{46}} +\frac{1}{X_{14}X_{47}X_{46}} +\frac{1}{X_{36}X_{16}X_{35}}+\frac{1}{X_{36}X_{37}X_{46}} +\frac{1}{X_{37}X_{47}X_{46}}\cr
&+\frac{1}{X_{14}X_{47}X_{57}}+\frac{1}{X_{36}X_{37}X_{35}}+\frac{1}{X_{37}X_{47}X_{57}} +\frac{1}{X_{14}X_{15}X_{16}}+\frac{1}{X_{14}X_{15}X_{57}}+\frac{1}{X_{15}X_{16}X_{35}}\cr
& +\frac{1}{X_{15}X_{35}X_{57}}+\frac{1}{X_{35}X_{37}X_{57}} \Bigg)
\end{split}
\end{equation}%

\section{Projective triangulations}
\label{triangulations}
In this section we study the projective triangulation of accordiohedra and relate it to the explicit results for scattering amplitudes obtained from recursion relations in Section \ref{Examples}.
In the context of $\phi^3$ theories the projective triangulation of associahedra was studied in \cite{CausalDiamonds,Yang}. 


Projective triangulation of a polytope is obtained by projecting a co-dimension $1$ facet onto a non-neighbouring co-dimension $k$ facet. For convenience we will stick to three dimensional accordiohedra, although our analysis holds for accordiohedra of all dimensions.
In the projective triangulation of three dimensional accordiohedra we project a face onto a reference edge labelled by two planar variables say $X_{ab}$ and $X_{cd}$.  
When we project a face (given by $X_{ij}=0$) whose edges are all either parallel or perpendicular to the reference edge, it gives rise to a prism with $X_{ij}$ as one of its faces. When the projected face is a quadrilateral this gives rise to triangular prisms. In all our examples, the projected face is a quadrilateral. However our analysis can be generalised to other polygonal faces as well. One may note that the spurious poles that appears in the recursion terms correspond to spurious boundaries within the accordiohedron formed upon triangulation.

Faces with one or more edges not parallel or perpendicular to the reference edge, when projected onto the reference edge give rise to positive geometries with curvy surfaces. In what follows we will relate the volume of triangular prisms thus obtained to the results of scattering amplitudes obtained by recursion relations in Section \ref{Examples}. For example, the number of non-adjacent faces to the reference edge in the accordiohedron will be the number of terms in the expression for the amplitude obtained via recursion.

The simple poles in the recursion term label the faces of the prism. For the explicit examples that we study in this paper it is a general feature that two of the poles in the corresponding partial amplitude are of the form $X_{kl}$ and $(\text{constant}-X_{kl})$ where $X_{kl}$ is one of the planar variables. These form the two parallel triangular faces of the prism. 

The prism is triangulated the following way. Take any vertex of the reference edge as the origin, say $Z_{\star}$. Connect it to any one of the vertices of the face ($X_{ij}=0$). Let us denote this vertex by $Z^{ij}_{m}$. Next we perform a full triangulation of the $X_{ij}$ face by connecting $Z_{m}^{ij}$ to its diagonally opposite vertex in $X_{ij}$, say $Z^{ij}_{n}$, which results in two triangles. The vertices of each triangle together with the origin $Z_{\star}$ form the four vertices of a simplex. The two simplices form the usual triangulation of the $X_{ij}$ face onto the reference vertex $Z_{\star}$. The remaining volume of the prism also forms a simplex. Thus the entire prism is triangulated via these three simplices and the total canonical function is obtained by adding the canonical functions of these simplices.

A novel feature that appears when we scale $1\le k\le d$ planar variables (where $d$ is the dimension of the accordiohedron)  is that some of the terms in the recursion can have poles that are quadratic in the planar variables. These correspond to  curvy surfaces inside the accordiohedron \cite{Yang}. This implies that the projection of the relevant facet $X_{ij}$ onto the reference edge leads to a curvy triangulation of the polytope. The curvy nature of the triangulation reflects the fact that some of the edges of the specific facet are neither parallel nor perpendicular to the reference edge. 

We end this section by giving the formula for the canonical function of a simplex whose vertices are labelled $Z_0$, $Z_1,\ldots, Z_m$  :
\begin{align}
\label{simplexformula}
[Z_0,Z_1,\ldots,Z_{m}]=\frac{\langle Z_0\,Z_1\,\ldots\,Z_{m}\rangle^{m}}{\prod_{i=0}^n\,\langle Y\,Z_0\ldots Z'_i\ldots Z_{m}\rangle}\,,
\end{align}
where angular bracket denotes determinant of the matrix formed by rows given by $Z_i$ , prime denotes omission of the corresponding vertex, and $Y=(1,\bf{X})$ where $\bf X$ denotes the planar variables labelling the origin.

\subsection{Quartic interaction}
We will now relate the results obtained via recursion relations  in Section \ref{n=10 amplitude} for the partial amplitudes of the four Stokes polytopes in the $n=10$ case to projective triangulation of the polytopes. Although recursion relations for the $\phi^4$ amplitudes were studied in \cite{Ryota} for the $n=8$ case, projective triangulation were not studied in that paper. We fill this gap here.
\subsection*{Cube}
Here we look at the case where the Stokes polytope is a cube. This corresponds to the case when the reference quadrangulation of the decagon is $Q=\{14,(5,10),69\}$. The equations that give the embedding of the accordiohedron in the kinematic space are given in \eqref{n=10cuberelations}.
We consider the scaling of $X_{14}$ and $X_{(5,10)}$. We saw in Section \ref{cube} that the partial amplitude from the recurrence relation has two terms and that their sum reproduces the correct partial amplitude.

Let us now  look at the second term in the expression for the partial amplitude \eqref{cubeeqn}. It takes the form :
\begin{equation}
\label{m49cube}
M_{49}= \frac{\epsilon_{(3,10)} ~ \epsilon_{49} \epsilon_{58} }{X_{14} (\epsilon_{49} - X_{(5,10)}) ~B~ X_{58} ~ X_{69}}
\end{equation}
where $B := \epsilon_{(3,10)} X_{(5,10)} - \epsilon_{49} X_{14}$ is the spurious pole that appears in this term.
Note that all the poles are linear in the planar variables. We will now identify this term as the canonical function of the prism obtained by projecting the $X_{49}$ facet onto the $X_{14}X_{(5,10)}$ line. Following the discussion at the beginning of this section we identify the prism to be as in Figure \ref{fig:prismcube}.
\begin{center}
\begin{figure}[h]
\begin{tikzpicture}
 \coordinate(Z1) at (0,0);
\node at (Z1) [left = 1mm of Z1] {$Z_1$};
\coordinate(Z2) at (5,0.7);
\node at (Z2) [right = .5mm of Z2] {$Z_2$};
\coordinate(Z*) at (1,-1.5);
\node at (Z*) [below = 1mm of Z*] {$Z_*$};
\coordinate(Z5) at (6.5,-.8);
\node at (Z5) [right = 1mm of Z5] {$Z_5$};
\coordinate(Z3) at (4.55,5.2);
\node at (Z3) [right= 1mm of Z3] {$Z_3$};
\coordinate(Z4) at (6.2,4.2);
\node at (Z4) [right=1mm of Z4] {$Z_4$};
\draw (Z1) -- (Z2);
\draw (Z*) -- (Z5);
\draw[newgreen]  (Z1) -- (Z*) node [midway,label = left:\textcolor{newgreen}{$X_{14},X_{(5,10)}$}] {};
\draw (Z2) -- (Z5);
\draw (Z2) -- (Z3);
\draw (Z4) -- (Z5);
\draw (Z4) -- (Z3);
\draw (Z4) -- (Z*);
\draw (Z1) -- (Z3);
\path[draw, fill=newgray, opacity =.15] (0,0) -- (5,0.7) -- (6.5,-.8) -- (1,-1.5) -- (0,0) -- cycle;
\node(X14) [inner sep = 0pt, newgray] at (3.3,-.4) {$X_{14}$};
\path[draw, fill=newpurple, opacity =.15] (5,0.7) -- (4.55,5.2) -- (6.2,4.2) -- (6.5,-.8) -- (5,0.7) -- cycle;
\node(X49) [inner sep = 0pt,newpurple] at (5.6,2.5) {$X_{49}$};
\path[draw, fill=newpink, opacity =.15] (0,0) -- (5,0.7) -- (4.55,5.2) -- (0,0)-- cycle;
\node(X69) [inner sep = 0pt, newpink] at (3.2,1.8) {$X_{69}$};
\node(B) [inner sep = 0pt, label = left:{$B = (\epsilon_{(3,10)}X_{(5,10)} - \epsilon_{49}X_{14})$}] at (1.5, 4.5) {};
\draw[->] (4.2,3.3) to[bend right] (B);
\node(X58) [inner sep = 0pt] at (6, -2) {$X_{58}$};
\draw[->] (4.5,0) to[bend left] (X58);
\end{tikzpicture}
\caption{\label{fig:prismcube}Prism formed by projecting $X_{49}$ facet onto $X_{14}X_{(5,10)}$ edge}
\end{figure}
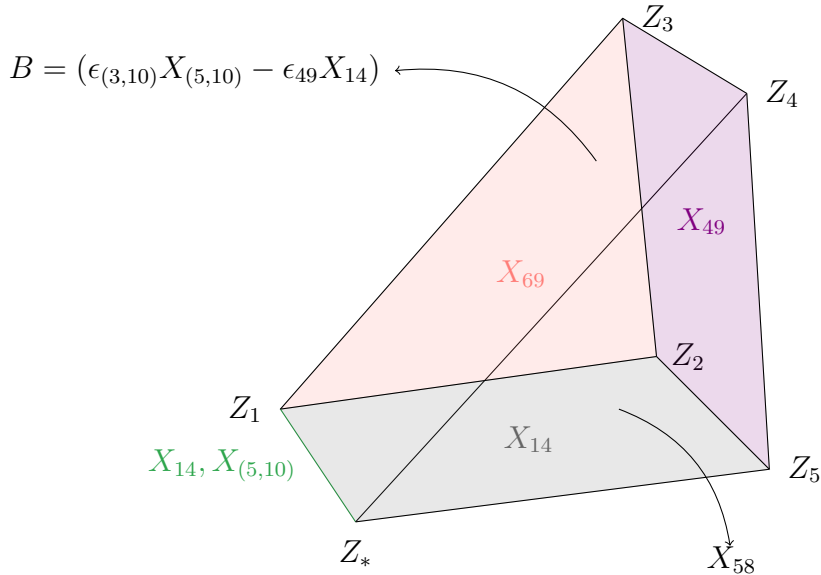
\end{center}
The vertices of the prism are given by :
 \begin{align}
 Z_{*}&\sim\{X_{14},X_{(5,10)},X_{58}\}=(1,0,0,0)\cr
Z_1&\sim\{X_{14},X_{(5,10)},X_{69}\}=(1,0,0,\epsilon_{58})\cr
Z_2&\sim\{X_{14},X_{69},X_{49}\}=(1,0,\epsilon_{49},\epsilon_{58})\cr
Z_3&\sim\{X_{49},X_{69},B\}=(1,\epsilon_{(3,10)},\epsilon_{49},\epsilon_{58})\cr
Z_4&\sim\{X_{49},X_{58},B\}=(1,\epsilon_{(3,10)},\epsilon_{49},0)\cr
Z_5&\sim\{X_{14},X_{49},X_{58}\}=(1,0,\epsilon_{49},0)
\end{align}
The canonical function of the prism is formed by adding the canonical function of the three simplices with vertices $\{Z_* Z_1 Z_2 Z_3\}$, $\{Z_* Z_2 Z_5 Z_4\}$ and $\{Z_* Z_2 Z_4 Z_3\}$  :
\begin{equation}
\Omega_{49}=[Z_* Z_1 Z_2 Z_3]+[Z_* Z_2 Z_5 Z_4]+[Z_* Z_2 Z_4 Z_3]
\end{equation}
Using \eqref{simplexformula} one can easily compute the above and see that the result matches the second term in the partial amplitude obtained in \eqref{cubeeqn}, namely $M_{49}$ as given in \eqref{m49cube}. In a similar fashion, the first term in the \eqref{cubeeqn} can be seen to correspond to the projective triangulation of the $X_{(3,10)}$ facet onto $X_{14}X_{(5,10)}$ edge.
%

\subsubsection*{Snake}
Let us now look at the snake case where the reference quadrangulation is $Q=\{14,16,18\}$. The equations that give the embedding of this accordiohedron in the kinematic space are given in \eqref{snake_constraints}.
There are a total of five terms in the partial amplitude obtained by recursion and their sum reproduces the correct result as shown in Section \ref{snake}.

\noindent Let us now  look at the fourth term in the recursion \eqref{snakeeqn}:
\begin{equation}
\label{m510snake}
M_{(5,10)}= \frac{\epsilon_{(3,10)} ~ \epsilon_{(5,10)} (\epsilon_{(5,10)} - \epsilon_{58} - \epsilon_{(7,10)})}{X_{14} ~X_{(5,10)} ~A~ B~ X_{710}}
\end{equation}
where $A :=\epsilon_{(5,10)} - \epsilon_{58} - X_{18},~B:= -\epsilon_{(5,10)} X_{14} + \epsilon_{(3,10)} X_{16} $ are the two spurious poles. All the poles are linear in the planar variables. The prism whose canonical function matches this term is obtained by projecting the $X_{(5,10)}$ face onto the $X_{14}X_{16}$ edge as given in Figure \ref{fig:snakeprism} : 
\begin{center}
\begin{figure}[h]
\begin{tikzpicture}
 \coordinate(Z1) at (0,0);
\node at (Z1) [left = 1mm of Z1] {$Z_1$};
\coordinate(Z2) at (5,0.7);
\node at (Z2) [right = .5mm of Z2] {$Z_2$};
\coordinate(Z*) at (1,-1.5);
\node at (Z*) [below = 1mm of Z*] {$Z_*$};
\coordinate(Z5) at (6.5,-.8);
\node at (Z5) [right = 1mm of Z5] {$Z_5$};
\coordinate(Z3) at (4.55,5.2);
\node at (Z3) [right= 1mm of Z3] {$Z_3$};
\coordinate(Z4) at (6.2,4.2);
\node at (Z4) [right=1mm of Z4] {$Z_4$};
\draw (Z1) -- (Z2);
\draw (Z*) -- (Z5);
\draw[newgreen]  (Z1) -- (Z*) node [midway,label = left:\textcolor{newgreen}{$X_{14},X_{16}$}] {};
\draw (Z2) -- (Z5);
\draw (Z2) -- (Z3);
\draw (Z4) -- (Z5);
\draw (Z4) -- (Z3);
\draw (Z4) -- (Z*);
\draw (Z1) -- (Z3);
\path[draw, fill=newgray, opacity =.15] (0,0) -- (5,0.7) -- (6.5,-.8) -- (1,-1.5) -- (0,0) -- cycle;
\node(X14) [inner sep = 0pt, newgray] at (3.3,-.4) {$X_{14}$};
\path[draw, fill=brown, opacity =.15] (5,0.7) -- (4.55,5.2) -- (6.2,4.2) -- (6.5,-.8) -- (5,0.7) -- cycle;
\node(X510) [inner sep = 0pt,brown] at (5.6,2.5) {$X_{(5,10)}$};
\path[draw, fill=newteal, opacity =.15] (0,0) -- (5,0.7) -- (4.55,5.2) -- (0,0)-- cycle;
\node(X710) [inner sep = 0pt, newteal] at (3.2,1.8) {$X_{(7,10)}$};
\node(B) [inner sep = 0pt, label = left:{$B = -\epsilon_{(5,10)}X_{14} + \epsilon_{(3,10)}X_{16}$}] at (1.5, 4.5) {};
\draw[->] (4.2,3.3) to[bend right] (B);
\node(A) [inner sep = 0pt] at (6, -2) {$A = \epsilon_{(5,10)} - \epsilon_{58} - X_{18}$};
\draw[->] (4.5,0) to[bend left] (A);
\end{tikzpicture}
\caption{\label{fig:snakeprism}Prism formed by projecting $X_{(5,10)}$ facet onto $X_{14}X_{16}$ edge.}
\end{figure}
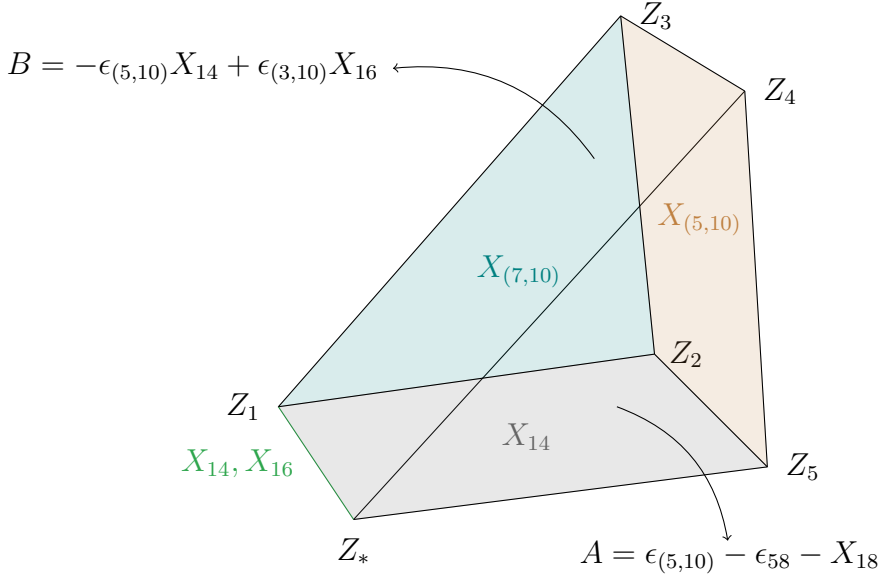
\end{center}
The vertices of this prism are given by : 
 \begin{align*}
 Z_{*}&\sim\{X_{14},X_{16},A\}=(1,0,0,0)\cr
Z_1&\sim\{X_{14},X_{16},X_{(7,10)}\}=(1,0,0,\epsilon_{(5,10)}-\epsilon_{58}-\epsilon_{(7,10)})\cr
Z_2&\sim\{X_{14},X_{(7,10)},X_{(5,10)}\}=(1,0,\epsilon_{(5,10)},\epsilon_{(5,10)}-\epsilon_{58}-\epsilon_{(7,10)})\cr
Z_3&\sim\{B,X_{(7,10)},X_{(5,10)}\}=(1,\epsilon_{(3,10)},\epsilon_{(5,10)},\epsilon_{(5,10)}-\epsilon_{58}-\epsilon_{(7,10)})\cr
Z_4&\sim\{A,B,X_{(5,10)}\}=(1,\epsilon_{(3,10)},\epsilon_{(5,10)},0)\cr
Z_5&\sim\{X_{14},A,X_{(5,10)}\}=(1,0,\epsilon_{(5,10)},0)
\end{align*}
%
The canonical function of the prism is the sum of functions of three simplices :
 \begin{equation}
\Omega_{(5,10)}=[Z_* Z_1 Z_2 Z_3]+[Z_* Z_2 Z_5 Z_4]+[Z_* Z_2 Z_4 Z_3]
\end{equation}
One can easily check that with $Y=(1,X_{14},X_{16},A)$, the canonical function $\Omega_{(5,10)}$ matches the term in the partial amplitude $M_{(5,10)}$ in \eqref{m510snake}. 

Let us now look at another term in the recursion \eqref{snakeeqn} given by :
\begin{equation}
M_{(3,10)}=\frac{C_{(3,10)}}{D_{(3,10)}}
\end{equation}
where  
\begin{align*}
D_{(3,10)}=(-\epsilon_{36} X_{14} &+ \epsilon_{(3,10)} (X_{14} - X_{16})) (\epsilon_{(5,10)} X_{14} - \epsilon_{(3,10)} X_{16}) (\epsilon_{(3,10)} - \epsilon_{38}- X_{18}) \cr
& (\epsilon_{58} X_{14} - \epsilon_{(3,10)} X_{16} + X_{14} X_{18}) X_{(3,10)} X_{(7,10)}
\end{align*}
and $C_{(3,10)}$ is combination of the constants $\epsilon_i$'s.

The pole $(\epsilon_{58} X_{14} - \epsilon_{(3,10)} X_{16} + X_{14} X_{18})$ is quadratic in the planar variables and corresponds to a curvy surface within the polytope. The other four terms in the recursion also correspond to ``Curvy Triangulation" due to the quadratic nature of the poles.


\subsection*{Lucas}
We will now look at the Lucas case where the reference quadrangulation is $Q=\{14,47,(7,10)\}$. The equations that give the embedding of the accordiohedron in the kinematic space are given in \eqref{lucas_constraints}.
We consider the scaling of $X_{14}$ and $X_{47}$. There are four terms in the corresponding recursion and their sum reproduces the correct partial amplitude as shown in Section \ref{lucas}.

Let us  look specifically at the first term in this recursion \eqref{lucaseqn} :
\begin{equation}
\label{m16lucas}
M_{16}= \frac{ \epsilon_{16} (\epsilon_{16} - \epsilon_{36}) \epsilon_{69}}{X_{14} ~X_{16} ~B ~X_{69}~X_{(7,10)}}
\end{equation}
where $B := \epsilon_{36}(X_{14}-X_{47})+\epsilon_{16}X_{47}$ is the spurious pole that appear in this term.
Note that all the poles in this term are linear in $X$. This term matches the canonical function of the prism obtained by projecting the $X_{16}$ facet onto the $X_{14}X_{47}$ edge as in Figure \ref{fig:lucasprism}
\begin{center}
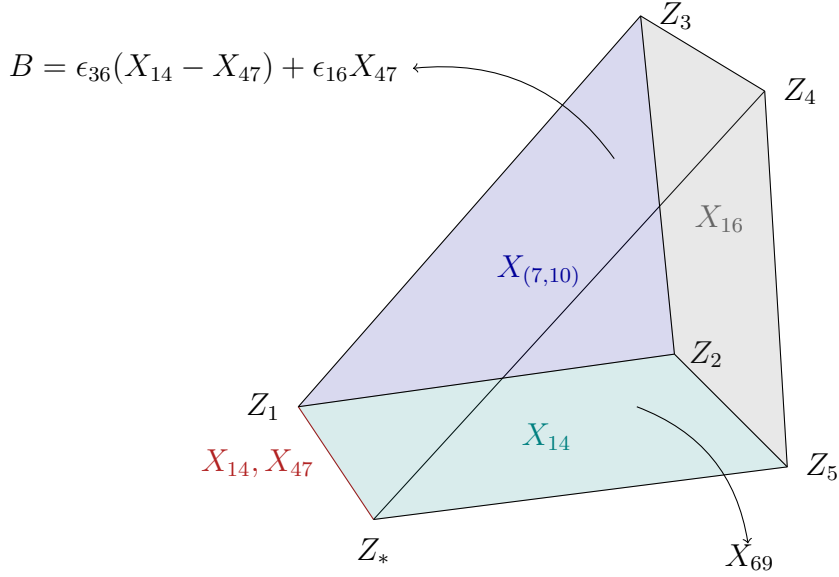
\begin{figure}[h]
\begin{tikzpicture}
\coordinate(Z1) at (0,0);
\node at (Z1) [left = 1mm of Z1] {$Z_1$};
\coordinate(Z2) at (5,0.7);
\node at (Z2) [right = .5mm of Z2] {$Z_2$};
\coordinate(Z*) at (1,-1.5);
\node at (Z*) [below = 1mm of Z*] {$Z_*$};
\coordinate(Z5) at (6.5,-.8);
\node at (Z5) [right = 1mm of Z5] {$Z_5$};
\coordinate(Z3) at (4.55,5.2);
\node at (Z3) [right= 1mm of Z3] {$Z_3$};
\coordinate(Z4) at (6.2,4.2);
\node at (Z4) [right=1mm of Z4] {$Z_4$};
\draw (Z1) -- (Z2);
\draw (Z*) -- (Z5);
\draw[newred]  (Z1) -- (Z*) node [midway,label = left:\textcolor{newred}{$X_{14},X_{47}$}] {};
\draw (Z2) -- (Z5);
\draw (Z2) -- (Z3);
\draw (Z4) -- (Z5);
\draw (Z4) -- (Z3);
\draw (Z4) -- (Z*);
\draw (Z1) -- (Z3);
\path[draw, fill=newteal, opacity =.15] (0,0) -- (5,0.7) -- (6.5,-.8) -- (1,-1.5) -- (0,0) -- cycle;
\node(X14) [inner sep = 0pt, newteal] at (3.3,-.4) {$X_{14}$};
\path[draw, fill=newgray, opacity =.15] (5,0.7) -- (4.55,5.2) -- (6.2,4.2) -- (6.5,-.8) -- (5,0.7) -- cycle;
\node(X16) [inner sep = 0pt,newgray] at (5.6,2.5) {$X_{16}$};
\path[draw, fill=newblue, opacity =.15] (0,0) -- (5,0.7) -- (4.55,5.2) -- (0,0)-- cycle;
\node(X710) [inner sep = 0pt, newblue] at (3.2,1.8) {$X_{(7,10)}$};
\node(B) [inner sep = 0pt, label = left:{$B = \epsilon_{36}(X_{14} - X_{47}) + \epsilon_{16}X_{47}$}] at (1.5, 4.5) {};
\draw[->] (4.2,3.3) to[bend right] (B);
\node(X69) [inner sep = 0pt] at (6, -2) {$X_{69}$};
\draw[->] (4.5,0) to[bend left] (X69);
\end{tikzpicture}
\caption{\label{fig:lucasprism}Prism formed by projecting $X_{16}$ facet onto $X_{14}X_{47}$ edge.}
\end{figure}
\end{center}
Let us now list the vertices of the prism : 
  \begin{align*}
 Z_{*}&\sim\{X_{14},X_{47},X_{69}\}=(1,0,0,0)\cr
Z_1&\sim\{X_{14},X_{47},X_{(7,10)}\}=(1,0,0,\epsilon_{69})\cr
Z_2&\sim\{X_{14},X_{16},X_{(7,10)}\}=(1,0,\epsilon_{16},\epsilon_{69})\cr
Z_3&\sim\{X_{16},X_{(7,10)},B\}=(1,\epsilon_{36}-\epsilon_{16},\epsilon_{36},\epsilon_{69})\cr
Z_4&\sim\{X_{16},X_{69},B\}=(1,\epsilon_{36}-\epsilon_{16},\epsilon_{36},0)\cr
Z_5&\sim\{X_{14},X_{16},X_{69}\}=(1,0,\epsilon_{16},0)
\end{align*}
This prism is formed by adding up three simplices :
\begin{equation}
\Omega_{16}=[Z_* Z_1 Z_2 Z_3]+[Z_* Z_2 Z_5 Z_4]+[Z_* Z_2 Z_4 Z_3]
\end{equation}
It can be easily checked that with $Y=(1,X_{14},X_{47},X_{69})$, the canonical function $\Omega_{16}$ matches $M_{16}$ in \eqref{m16lucas}.

Now lets look into the fourth term in the recursion \eqref{snakeeqn},
\begin{equation}
M_{49}= \frac{\epsilon_{(3,10)}(\epsilon_{69}-\epsilon_{49})}{X_{14}~ X_{49} ~X_{69} (\epsilon_{49}X_{14}+\epsilon_{(3,10)}X_{47}-X_{14}X_{(7,10)})}
\end{equation}
Note that the pole $(\epsilon_{49}X_{14}+\epsilon_{(3,10)}X_{47}-X_{14}X_{(7,10)})$ is quadratic in the planar variables and corresponds to a curvy surface within polytope. This corresponds to the projective triangulation of the $X_{49}$ facet onto the $X_{14}X_{47}$ line. 

\subsection*{Mixed type}
Here we look into the mixed case where the reference quadrangulation is $Q=\{14,(5,10),69\}$. The equations that give the embedding of the accordiohedron in the kinematic space are given in \eqref{n=10mixedrelations}.
We consider the scaling of $X_{14}$ and $X_{(5,10)}$. There are three terms in the corresponding recursion and their sum reproduces the correct partial amplitude as shown in Section \ref{mixed}.

Let us now  look at the third term in the recursion \eqref{mixedeqn}:
\begin{equation}
\label{M49mixed}
M_{49}= \frac{\epsilon_{(3,10)} ~ \epsilon_{49} (\epsilon_{47} -\epsilon_{49}+\epsilon_{69})}{X_{14} ~X_{49} ~A~ B ~ X_{69}}
\end{equation}
where $A:=-\epsilon_{47}+\epsilon_{49}-X_{(7,10)},~B := \epsilon_{49} X_{14}-\epsilon_{(3,10)} X_{(5,10)}  $ are the two spurious poles that appear in this term.
Note that all the poles are linear in $X$. This term matches the canonical function of the prism in Figure \ref{fig:mixedprism} obtained by projecting the face $X_{49}$ onto the $X_{14}X_{(5,10)}$ edge.
Let us now list down the vertices of this prism : 
 \begin{align*}
 Z_{*}&\sim\{X_{14},X_{(5,10)},A\}=(1,0,0,0)\cr
  Z_1&\sim\{X_{14},X_{(5,10)},X_{69}\}=(1,0,0,-\epsilon_{47}+\epsilon_{49}-\epsilon_{69})\cr
Z_2&\sim\{X_{14},X_{69},X_{49}\}=(1,0,\epsilon_{49},-\epsilon_{47}+\epsilon_{49}-\epsilon_{69})\cr
Z_3&\sim\{X_{49},X_{69},B\}=(1,\epsilon_{(3,10)},\epsilon_{49},-\epsilon_{47}+\epsilon_{49}-\epsilon_{69})\cr
Z_4&\sim\{X_{49},A,B\}=(1,\epsilon_{(3,10)},\epsilon_{49},0)\cr
Z_5&\sim\{X_{14},X_{49},A\}=(1,0,\epsilon_{49},0)
\end{align*}
\begin{center}
\begin{figure}[H]
\begin{tikzpicture}
\coordinate(Z1) at (0,0);
\node at (Z1) [left = 1mm of Z1] {$Z_1$};
\coordinate(Z2) at (5,0.7);
\node at (Z2) [right = .5mm of Z2] {$Z_2$};
\coordinate(Z*) at (1,-1.5);
\node at (Z*) [below = 1mm of Z*] {$Z_*$};
\coordinate(Z5) at (6.5,-.8);
\node at (Z5) [right = 1mm of Z5] {$Z_5$};
\coordinate(Z3) at (4.55,5.2);
\node at (Z3) [right= 1mm of Z3] {$Z_3$};
\coordinate(Z4) at (6.2,4.2);
\node at (Z4) [right=1mm of Z4] {$Z_4$};
\draw (Z1) -- (Z2);
\draw (Z*) -- (Z5);
\draw[newgreen]  (Z1) -- (Z*) node [midway,label = left:\textcolor{newgreen}{$X_{14},X_{(5,10)}$}] {};
\draw (Z2) -- (Z5);
\draw (Z2) -- (Z3);
\draw (Z4) -- (Z5);
\draw (Z4) -- (Z3);
\draw (Z4) -- (Z*);
\draw (Z1) -- (Z3);
\path[draw, fill=newgray, opacity =.15] (0,0) -- (5,0.7) -- (6.5,-.8) -- (1,-1.5) -- (0,0) -- cycle;
\node(X14) [inner sep = 0pt, newgray] at (3.3,-.4) {$X_{14}$};
\path[draw, fill=newblue, opacity =.15] (5,0.7) -- (4.55,5.2) -- (6.2,4.2) -- (6.5,-.8) -- (5,0.7) -- cycle;
\node(X49) [inner sep = 0pt,newblue] at (5.6,2.5) {$X_{49}$};
\path[draw, fill=newred, opacity =.15] (0,0) -- (5,0.7) -- (4.55,5.2) -- (0,0)-- cycle;
\node(X69) [inner sep = 0pt, newred] at (3.2,1.8) {$X_{69}$};
\node(B) [inner sep = 0pt, label = left:{$B = \epsilon_{49}X_{14} - \epsilon_{(3,10)}X_{(5,10)}$}] at (1.5, 4.5) {};
\draw[->] (4.2,3.3) to[bend right] (B);
\node(A) [inner sep = 0pt] at (6, -2) {$A = -\epsilon_{47} + \epsilon_{49} - X_{(7,10)}$};
\draw[->] (4.5,0) to[bend left] (A);
\end{tikzpicture}
\caption{\label{fig:mixedprism}Prism formed by projecting $X_{49}$ facet onto $X_{14}X_{(5,10)}$ edge.}
\end{figure}
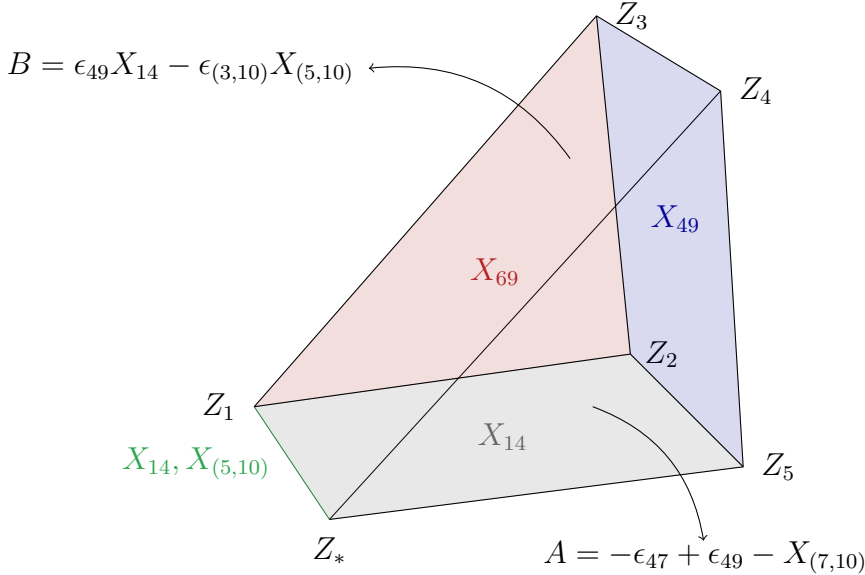
\end{center}
The canonical function of the prism is given by the sum of the functions of three simplices :
\begin{equation}
\Omega_{49}=[Z_* Z_1 Z_2 Z_3]+[Z_* Z_2 Z_5 Z_4]+[Z_* Z_2 Z_4 Z_3]
\end{equation}
%
%
%
It can be easily checked that with $Y=(1,X_{14},X_{(5,10)},A)$, the canonical function $\Omega_{49}$ matches $M_{49}$ \eqref{M49mixed}.
Now let us look at the second term in the recursion \eqref{mixedeqn},
\begin{equation}
M_{47}= \frac{\epsilon_{(3,10)}(\epsilon_{49}-\epsilon_{47})(\epsilon_{47}+X_{(7,10)})}{(X_{14}~ X_{47} ~X_{(7,10)})(\epsilon_{47}-\epsilon_{49}+X_{(7,10)}) (\epsilon_{47}X_{14}-\epsilon_{(3,10)}X_{(5,10)}+X_{14}X_{(7,10)})}
\end{equation}
The pole $(\epsilon_{47}X_{14}-\epsilon_{(3,10)}X_{(5,10)}+X_{14}X_{(7,10)})$ is quadratic in planar variables and corresponds to a curvy surface within polytope. This corresponds to projective triangulation of the $X_{47}$ face onto the $X_{14}X_{(5,10)}$ line. 

\subsection{Polynomial interactions}
We will now consider theories with mixed vertices (or polynomial interactions) and show that the same type of projective triangulation that we discussed above appears where each term in the recursion corresponds to a specific projective triangulation of the accordiohedron. 

\subsubsection*{$n=7$ in $\phi^3+\phi^4$}
Here we look into the case where the reference quadrangulation is $Q=\{14,16,46\}$ discussed in Section \ref{poly_int}. The equations that give the embedding of the accordiohedron in the kinematic space are given in \eqref{n=7mixedrelations}.
We consider the scaling of $X_{14}$ and $X_{16}$. There are a total of five terms in the corresponding recursion and their sum correctly reproduces the correct amplitude.
Let us now look at the third term in the recursion \eqref{n=7eqn} :
\begin{equation}
A_{36}= \frac{\epsilon_{35} ~ \epsilon_{36} (\epsilon_{36} -\epsilon_{37})}{X_{16} ~X_{36} ~A~ X_{35} ~ X_{46}}
\end{equation}
where $B := \epsilon_{36}X_{14}+\epsilon_{37}(-X_{14}+X_{16})$ is the spurious pole that appears in this term.
All the poles are linear in the planar variables. The prism in Fig \ref{fig:n=7accordiohedronprism} whose canonical function matches this term is obtained by projecting the $X_{36}$ facet onto the $X_{14}X_{16}$ line. 
\begin{center}
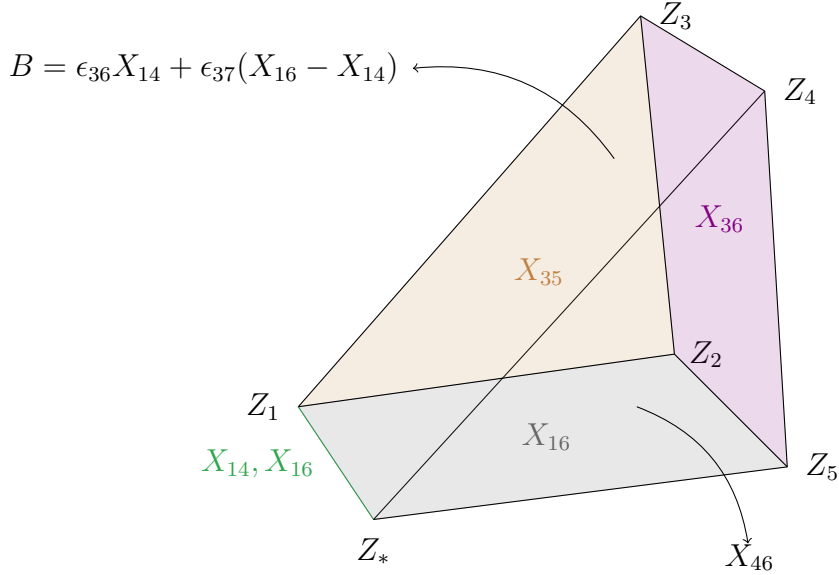
\begin{figure}[h]
\begin{tikzpicture}
\coordinate(Z1) at (0,0);
\node at (Z1) [left = 1mm of Z1] {$Z_1$};
\coordinate(Z2) at (5,0.7);
\node at (Z2) [right = .5mm of Z2] {$Z_2$};
\coordinate(Z*) at (1,-1.5);
\node at (Z*) [below = 1mm of Z*] {$Z_*$};
\coordinate(Z5) at (6.5,-.8);
\node at (Z5) [right = 1mm of Z5] {$Z_5$};
\coordinate(Z3) at (4.55,5.2);
\node at (Z3) [right= 1mm of Z3] {$Z_3$};
\coordinate(Z4) at (6.2,4.2);
\node at (Z4) [right=1mm of Z4] {$Z_4$};
\draw (Z1) -- (Z2);
\draw (Z*) -- (Z5);
\draw[newgreen]  (Z1) -- (Z*) node [midway,label = left:\textcolor{newgreen}{$X_{14},X_{16}$}] {};
\draw (Z2) -- (Z5);
\draw (Z2) -- (Z3);
\draw (Z4) -- (Z5);
\draw (Z4) -- (Z3);
\draw (Z4) -- (Z*);
\draw (Z1) -- (Z3);
\path[draw, fill=newgray, opacity =.15] (0,0) -- (5,0.7) -- (6.5,-.8) -- (1,-1.5) -- (0,0) -- cycle;
\node(X16) [inner sep = 0pt, newgray] at (3.3,-.4) {$X_{16}$};
\path[draw, fill=newpurple, opacity =.15] (5,0.7) -- (4.55,5.2) -- (6.2,4.2) -- (6.5,-.8) -- (5,0.7) -- cycle;
\node(X36) [inner sep = 0pt,newpurple] at (5.6,2.5) {$X_{36}$};
\path[draw, fill=brown, opacity =.15] (0,0) -- (5,0.7) -- (4.55,5.2) -- (0,0)-- cycle;
\node(X35) [inner sep = 0pt, brown] at (3.2,1.8) {$X_{35}$};
\node(B) [inner sep = 0pt, label = left:{$B = \epsilon_{36}X_{14} +  \epsilon_{37}(X_{16} - X_{14})$}] at (1.5, 4.5) {};
\draw[->] (4.2,3.3) to[bend right] (B);
\node(X46) [inner sep = 0pt] at (6, -2) {$X_{46}$};
\draw[->] (4.5,0) to[bend left] (X46);
\end{tikzpicture}
\caption{\label{fig:n=7accordiohedronprism}Prism formed by projecting $X_{36}$ facet onto $X_{14}X_{16}$ edge.}
\end{figure}
\end{center}
To see this we will now list down the vertices of the prism : 
 \begin{align*}
Z_{*}&\sim\{X_{14},X_{16},X_{46}\}=(1,0,0,0)\cr
Z_1&\sim\{X_{14},X_{16},X_{35}\}=(1,0,0,\epsilon_{35})\cr
Z_2&\sim\{X_{16},X_{35},X_{36}\}=(1,\epsilon_{36},0,\epsilon_{35})\cr
Z_3&\sim\{X_{36},X_{35},B\}=(1,\epsilon_{37},\epsilon_{37}-\epsilon_{36},\epsilon_{35})\cr
Z_4&\sim\{X_{36},X_{46},B\}=(1,\epsilon_{37},\epsilon_{37}-\epsilon_{36},0)\cr
Z_5&\sim\{X_{16},X_{36},X_{46}\}=(1,\epsilon_{36},0,0)
\end{align*}
%
The canonical function of the prism is given by the following sum :
\begin{equation}
\Omega_{36}=[Z_* Z_1 Z_2 Z_3]+[Z_* Z_2 Z_5 Z_4]+[Z_* Z_2 Z_4 Z_3]
\end{equation}
It can be easily checked that with $Y=(1,X_{14},X_{16},X_{46})$, the canonical function $\Omega_{36}$ matches with amplitude $A_{36}$.
Let us now look at the second term in the recursion \eqref{n=7eqn}:
\begin{equation}
A_{15}=\frac{(\epsilon_{35}-\epsilon_{15})\epsilon_{57}}{X_{16}~A~X_{15}~X_{35}}
\end{equation}
where $A := \epsilon_{57}X_{14}+X_{16}(\epsilon_{15}-X_{46})$ is the spurious pole that appear in this term. This term has a pole in $X_{16}X_{46}$ variable and corresponds to a curvy surface within the polytope. It corresponds to a projective triangulation of the $X_{15}$ fact onto the $X_{14}X_{16}$ edge.
The other three terms in the recursion similarly correspond to the curvy triangulation of the accordiohedron.

\section{1-loop in $\phi^4$}
\label{loopsection}
In the proof of the recursion relations \eqref{generalscalingformula} we had emphasised that it goes through provided one has a linear relation among the kinematic variables and that the associated positive geometry is a simple polytope. In this section we illustrate this by applying the recursion relations to the one loop amplitude of the quartic scalar theory. We then relate these to projective triangulations of the polytope.


In \cite{CausalDiamonds} a geometric realisation of $D$-type associahedra was shown to reproduce the one-loop integrand in the cubic theory. A combinatorial realisation of the entire class of such polytopes was realised by pseudo-triangulations of a polygon with an annulus in its interior \cite{Ceballos}. 
Recently in \cite{alok1loop} this analysis was extended to the quartic theory  and it was shown that the one-loop integrand of the quartic theory is given by the canonical top-form of a class of polytopes called pseudo-accordiohedra realised  by suitable projections of $D$-type associahedra in an abstract kinematic space. The abstract kinematic space in which these are realised is an $n^2$ dimensional space spanned by the usual planar variables $X_{ij}$ along with additional variables which we label $X_{i\bar j}=X_{\bar i j}$, $Y_i$ and $\widetilde Y_i$. The $Y_i(\widetilde Y_i)$ are associated  with the momenta running in the loop just before the external leg $i$.  In the loop integrand, the $X_{i\bar j}$ variables are given by $X_{i\bar j}=X_{\bar i j}=(p_j+\ldots+p_n+p_1+\ldots+p_{i-1})^2$. We refer the reader to \cite{alok1loop} for details.

\begin{figure}[h]
\centering
\begin{tikzpicture}
\coordinate(1) at (0,3.5) {};
\coordinate(2) at (3,1) {};
\coordinate(3) at (2,-2) {};
\coordinate(4) at (-2,-2) {};
\coordinate(5) at (-3,1) {};
\draw (1)--(2) node[midway, label = right:{$X_{14}$}] {};
\draw (2)--(3) node[midway, label = right:{$\widetilde{Y}_4$}] {};
\draw (3)--(4) node[midway, label = below:{$X_{3\bar{4}}$}] {};
\draw (4)--(5) node[midway, label = left:{$Y_3$}] {};
\draw (5)--(1) node[midway, label = left:{$Y_1$}] {};
\end{tikzpicture}
\caption{\label{loopacco}Pseudo-accordiohedron $\mathcal{P AC}(Q_1)$ for $n = 4$}
\end{figure}
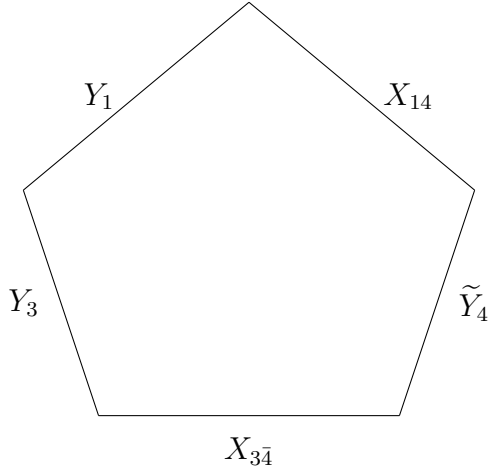

In the $n=4$ case, there are two primitive pseudo-accordiohedra and a weighted sum of their canonical functions reproduces the full one-loop integrand. We focus on the pseudo-accordiohedron $\mathcal{PAC}(Q_1)$ which is a pentagon whose faces are labelled by $X_{14}, Y_1, \widetilde Y_4, X_{3\bar 4}$ and  $Y_3$ as shown in Figure~\ref{loopacco}.

%
The embedding of this pseudo-accordiohedron in the abstract kinematic space is given by the following constraints \cite{alok1loop} :
\begin{align}
Y_1&\ge 0\nonumber\cr
X_{14}&\ge 0\nonumber\cr
Y_3&= Y_1-X_{14}+\epsilon_{Y_3}\ge 0\nonumber\cr
\widetilde Y_4&= -Y_1+\epsilon_{\bar Y_4}\ge 0\nonumber\cr
X_{3,\bar 4}&=-X_{14}+\epsilon_{X_{3,\bar 4}}\ge 0
\end{align}
Let us now consider the scaling of one of the basis variables, say $X_{14} \rightarrow z X_{14}$. From the above constraints, the variables that are deformed under this scaling are $(Y_3, X_{3,\bar 4})$. The recursion relation \eqref{singlescalingformula} takes the following form:
\begin{equation}
\label{loopsingle}
m_4(\mathcal Q_1)=\left(\frac{1}{Y_{3}}+\frac{1}{X_{14}}\right)\left(\frac{1}{X_{3\bar 4}-Y_3}+\frac{1}{Y_{1}}\right)+\left(\frac{1}{X_{3 \bar 4}}+\frac{1}{X_{14}}\right)\left(\frac{1}{\widetilde Y_{4}}+\frac{1}{Y_{3}-X_{3\bar 4}}\right)
\end{equation}
This matches the expected result \cite{alok1loop} for the partial amplitude associated to this pseudo-accordiohedron.
\begin{figure}[H]
\centering
\begin{tikzpicture}
\coordinate(1) at (0,0) {};
\coordinate(2) at (0,4) {};
\coordinate(3) at (4,4) {};
\coordinate(4) at (4,2) {};
\coordinate(5) at (2.5,0) {};
\draw (1)--(2) node[midway, label = left:\textcolor{newred}{$X_{14}$}] {};
\draw (2)--(3) node[midway, label = above:\textcolor{newred}{$\widetilde{Y}_4$}] {};
\draw (3)--(4) node[midway, label = right:\textcolor{newred}{$X_{3\bar{4}}$}] {};
\draw (4)--(5) node[midway, label = right:\textcolor{newred}{$Y_3$}] {};
\draw (5)--(1) node[midway, label = below:\textcolor{newred}{$Y_1$}] {};
\draw[newblue] (0,2) -- (4,2);
\end{tikzpicture}
\caption{\label{prismloop}Two-dimensional prisms formed by projecting the edges $Y_3$ and $X_{3\bar4}$ onto the $X_{14}$}
\end{figure}
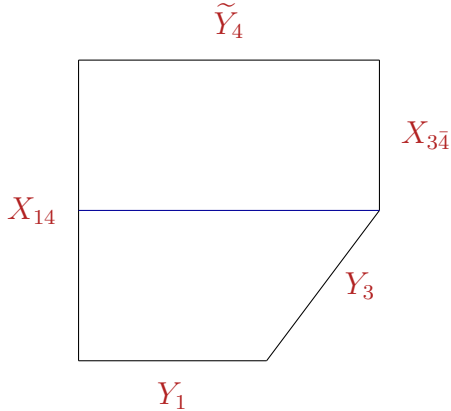%
The two terms in the above recursion relation \eqref{loopsingle} correspond to the canonical functions of the two dimensional-prisms obtained by projecting the edges $Y_3$ and $X_{3\bar4}$ respectively onto the $X_{14}$ edge as in Figure~\ref{prismloop}. This provides a complete triangulation of the pentagon. 

\section{Conclusion}
\label{conclusion}

In this paper, we investigated some aspects of the positive geometry of theories with polynomial interaction. In these theories, the planar amplitude is obtained from a weighted sum of all the canonical forms related to various accordiohedra. We have shown that, as in the case of $\phi^p$ interactions, all the weights can be determined using the factorization property of accordiohedra. Thus the full tree-level planar amplitude of these theories can be determined purely from the corresponding positive geometry. 

\par
In theories with polynomial interaction, there is an ambiguity for weights related to the same accordiohedron : for example, in the case of $\phi^3 + \phi^4$ theory with $n=6$ particles, the accordiohedra obtained from dissections 1 and 2 are equivalent under cyclic permutation. In such cases we cannot fix their weights uniquely. This ambiguity appears only in the case of polynomial interactions. In $\phi^p$ theories different dissections give rise to different accordiohedra. It will be interesting to explore the implication of these facts in field theory. 

\par

We derived recursion relations for the partial amplitudes in these theories following the derivation for cubic theories in \cite{Yang,Ryota}. In our derivation we made use of the canonical convex embedding of accordiohedra in kinematic space.
We emphasise that the derivation is very general in nature and is applicable to any scalar field theory whose positive geometry is a simple polytope and which has a linear relation as in \eqref{compatibleX} among the kinematic variables. In particular we expect them to hold true for loop amplitudes too if the corresponding positive geometry is a simple polytope and the planar variables satisfy similar linear relationships as in (4.4). We have illustrated this in the simple example of the one-loop integrand in the quartic theory for $n=4$. We leave a more detailed investigation for the future.

 \par
We showed that the results from the recursion relate to projective triangulations of accordiohedra. A novel feature that appears when we rescale $1< k < d$ planar variables (where $d$ is the dimension of the accordiohedron) is that some of the terms in the recursion have poles that are quadratic in the planar variables. This leads to a ``curvy triangulation" of the accordiohedron. It will be interesting to investigate the application of such triangulations and recursion relations to the $\mathcal{N}=4$ sYM and the amplituhedron. \par 
Though we have not discussed the triangulation corresponding to the recursion relation obtained via. the rescaling of all basis variables, it can be checked that the terms in the resulting recursion will correspond to the projection of non-neighbouring co-dimension one facets onto the reference vertex. Further we can see that interesting geometries appear from triangulations that correspond to recursion relations obtained by a single variable rescaling. Here one projects a non-neighbouring co-dimension one facet onto the reference co-dimension one facet labeled by the scaled basis variable. 
Recursion relations in this case have only simple poles and hence will correspond to flat surfaces inside the accordiohedron unlike curvy surfaces which are very special to the case of  scaling  $1< k < d$ basis variables.

\section*{Acknowledgments}
R.R.J and S.M thank Sujay K. Ashok, Mrunmay Jagadale and Alok Laddha for discussions on related topics.  R.K would like to thank Hiroyuki Ishida for insightful discussions. We thank Anupa Sunny for the figures in Section \ref{triangulations}. R.R.J thanks IISER Pune for hospitality where the work was done. The work of R.R.J is supported by the MIUR PRIN Contract 2015 MP2CX4 ``Non-perturbative Aspects Of Gauge Theories And Strings'' and is also partially supported by ``Fondi Ricerca Locale dell’Universit\`a del Piemonte Orientale''.  We thank all the COVID-19 workers for their commendable services. S.M. thanks his friends for keeping the spirit high during this difficult period.

\appendix

\section{$n=8$ in $\phi^3+\phi^4+\phi^5$}
\label{App}
In this Appendix we will consider another example of polynomial interaction namely $n=8$ amplitude with $\phi^3+\phi^4+\phi^5$ vertices to illustrate the validity of the recursion relation \eqref{generalscalingformula} and the triangulations that follow from it.   In this case there are nine accordiohedra as shown in Figure \ref{fig:345n8accordiohedron}). We consider the first case with reference dissection $\{13,48\}$ and the corresponding accordiohedron is given by a square. The compatible diagonals are $\{28,37\}$.The embedding of the polytope in the positive kinematic space is given by the following relations between the corresponding planar variables :
\begin{align}
\label{n=8mixedrelations}
X_{28}&=-X_{13}+\epsilon_{28}\cr
X_{37}&=-X_{48}+\epsilon_{48}
\end{align}
Consider first the single variable scaling $X_{13} \to z X_{13}$. The corresponding recursion relation gives the simple answer
\begin{equation}
\label{single_amp}
A=\left(\frac{1}{X_{13}}+\frac{1}{X_{28}}\right)\left(\frac{1}{X_{48}}+\frac{1}{X_{37}}\right)
\end{equation}
It can be easily checked that this matches the expected result:
\begin{align}
\label{amp_exp}
\widetilde A_8^{(13,48)} =\frac{1}{X_{13}X_{48}}+\frac{1}{X_{48}X_{28}}+\frac{1}{X_{28}X_{37}}+\frac{1}{X_{37}X_{13}}.
\end{align}

The recursion term \eqref{single_amp} corresponds to a complete triangulation of the accordiahedron obtained  by projecting the $X_{28}$ edge onto the $X_{13}$ edge. Consistent with the fact \eqref{single_amp} is the canonical function of this square accordiahedron \cite{Salvatori:2019phs}.\\

Next we consider a more non-trivial case of  all basis variable scaling,
\begin{equation}
X_{13} \to z X_{13}, \quad X_{48}\to z X_{48}
\end{equation}
The set of deformed variables is $\{X_{28},X_{37}\}$. The recursion formula \eqref{generalscalingformula} takes the form:
\begin{align}
\label{double_amp}
A_8^{(13,48)}&=\frac{\hat z_{28}^2}{X_{28}} \left( \frac{1}{\hat z_{28} X_{48}}+\frac{1}{-\hat z_{28} X_{48}+\epsilon_{37}}\right)+\frac{\hat z_{37}^2}{X_{37}}\left(\frac{1}{-\hat z_{37} X_{13}+\epsilon_{28}}+\frac{1}{\hat z_{37} X_{13}}\right)
\end{align}
where
\begin{equation}
\hat z_{28}=\frac{\epsilon_{28}}{X_{13}} , \quad \hat z_{37}=\frac{\epsilon_{37}}{X_{48}}
\end{equation}
By a straightforward computation it can be checked that this matches the expected result $\widetilde A_8^{(13,48)} $ \eqref{amp_exp}. The two terms in the recursion \eqref{double_amp}
 correspond to the canonical functions of two triangles obtained by projecting the $X_{28}$ and $X_{37}$ edges onto the $X_{13}X_{48}$ vertex respectively, thus providing another complete triangulation of the accordiahedron.

\bibliographystyle{JHEP}

\end{document}